\pdfoutput=1
\documentclass[]{interact}

\usepackage{epstopdf}% To incorporate .eps illustrations using PDFLaTeX, etc.
\usepackage[caption=false]{subfig}% Support for small, `sub' figures and tables
\usepackage[backend=biber,style=chem-acs,articletitle=true]{biblatex}
\let\cite=\supercite
\DeclareCiteCommand{\citenum}
  {}
  {\printfield{labelnumber}}
  {}
  {}
\renewcommand\bibfont{\fontsize{10}{12}\selectfont}% Bibliography support using natbib.sty

\theoremstyle{plain}% Theorem-like structures provided by amsthm.sty

\theoremstyle{definition}

\theoremstyle{remark}

\usepackage{float} 

\usepackage{physics}
\usepackage{braket}
\usepackage{mhchem}
\usepackage{color}

\usepackage{tabularx}
\usepackage{siunitx}
\usepackage{xcolor}

\begin{document}

%\articletype{ARTICLE TEMPLATE}% Specify the article type or omit as appropriate

\title{Optimization stability in excited state-specific variational Monte Carlo}

\author{
\name{Leon Otis\textsuperscript{a} and Eric Neuscamman\thanks{CONTACT Eric Neuscamman eneuscamman@berkeley.edu}\textsuperscript{b,}\textsuperscript{c}}
\affil{\textsuperscript{a}Department of Physics, University of California Berkeley, CA 94720,USA; \textsuperscript{b}Department of Chemistry, University of California Berkeley, CA 94720,USA;\textsuperscript{c}Chemical Sciences Division, Lawrence Berkeley National Laboratory, Berkeley, CA, 94720, USA}
}

\maketitle

\begin{abstract}
We investigate the issue of optimization stability in variance-based state-specific 
variational Monte Carlo, discussing the roles of the objective function, the complexity of wave
function ansatz, the amount of sampling effort, and the choice of minimization algorithm.
Using a small cyanine dye molecule as a test case, we systematically perform minimizations 
using variants of the linear method as both a standalone algorithm and in a hybrid combination 
with accelerated descent.
We demonstrate that adaptive step control is crucial for maintaining the linear method's 
stability when optimizing complicated wave functions and that the hybrid method enjoys 
both greater stability and minimization performance.

\end{abstract}


\section{Introduction}

The challenges of wave function optimization have been a long-running thread in the development
of Quantum Monte Carlo (QMC) methods.
Ground state variational Monte Carlo (VMC) alone has seen significant evolution both in 
the optimization algorithms themselves\cite{Nightingale2001,Casalegno2003,Umrigar2005,Sorella2001,Sorella2005,Toulouse2007,Umrigar2007,Sorella2007,Neuscamman2012,Schwarz2017,Sabzevari2018,Luo2019,Mahajan2019,Sabzevari2020,Otis2019} and the objective functions they are applied to.\cite{Umrigar1988,Kent1999,Lin2000,Foulkes2001,Umrigar2007}
Early work on VMC optimization implemented steepest descent\cite{Huang1996,Huang1999} and Newton-Raphson\cite{Lin2000,Lee2005} for 
relatively small numbers of parameters in atoms and small molecules.
As optimization algorithms, steepest descent and Newton-Raphson mark out two extremes in 
terms of derivative information and computational cost as purely first-order and fully 
second-order methods respectively.
Subsequent methods, including approximate Newton,\cite{Umrigar2005,Sorella2005} the linear method (LM),\cite{Nightingale2001,Umrigar2007,Toulouse2007,Sabzevari2020} stochastic reconfiguration,\cite{Sorella2001,Sorella2007,Neuscamman2012} and accelerated descent (AD),\cite{Schwarz2017,Sabzevari2018,Luo2019,Mahajan2019} can be broadly viewed as 
pursuing a balance between the advantages of second-order derivative information 
against the cost of obtaining it.
This profusion of different techniques, spread among many research groups, has 
motivated recent efforts to systematically compare their performance and identify 
effective combinations of them.\cite{Otis2019}

The need to achieve robust VMC optimization has also surfaced more recently in the context of excited state VMC.
For this frontier in VMC, both energy-based state-averaged and 
variance-based state-specific objective functions have been developed for describing excited states.
State-averaged energy minimization has been successful in many applications,\cite{Cordova2007,Filippi2009,Send2011,Guareschi2013,Guareschi2016,Cuzzocrea2020,Dash2021} while the state-specific approach\cite{Zhao2016,Shea2017,Flores2019,Garner2020,Otis2020,Otis2021} offers advantages in cases where 
the appropriate orbital shapes for the relevant states differ significantly.
These considerations are also present in deterministic methods such as 
CASPT2\cite{Domingo2012,Meyer2014,Tran2019} and are especially important when
studying charge transfer states\cite{Dutta2018,Kozma2020,Loos2021,Otis2021} and core excitations.\cite{Garner2020}
Recent work has sought to also achieve state-specificity in energy minimization by imposing orthogonality constraints.\cite{Pathak2021}

Methodological advancements have steadily occurred in recent years for both state-averaged 
and state-specific VMC.
Many innovations can be readily adopted within both frameworks such as improvements in ansatz 
design with selected CI (sCI) expansions\cite{Dash2018,Dash2019,Dash2021} and new optimization 
algorithms.\cite{Zhao2017,Otis2019,Otis2020,Otis2021}
In addition, multiple protocols\cite{Dash2019,Dash2021,Robinson2017,Garner2020,Otis2021} have 
been developed to ensure balanced treatments of different states, a prerequisite for obtaining 
accurate excitation energies.
The progress on all these fronts has helped advance excited-state VMC from initial 
applications\cite{Umrigar1988,Schautz2004,Umrigar2005,Zhao2016} with small molecules and simple wave functions to 
much wider use on larger systems with more sophisticated ansatzes.\cite{Otis2021,Dash2021}
 
Given the promise of VMC as a reliable tool for excited states, recent 
research\cite{Cuzzocrea2020} raising concerns about the stability of variance minimization in 
VMC merits close attention.
In a small cyanine dye molecule, Filippi and coworkers found failures of 
variance minimization to converge to the 
correct excited state even for high quality wave function ansatzes.
In some cases, a lowering of the variance was accompanied by a substantial rise in the energy 
as the optimization moved to some undesired state.
The authors attributed this unstable behavior to the shape of the variance optimization surface, with the 
optimization finding ``little or no barrier to escape from a local minimum or local 
plateau, eventually converging to a lower-variance state instead of the target state."\cite{Cuzzocrea2020}
This hypothesis would imply that the same behavior might be observed in other excited state 
applications of variance minimization, posing a significant obstacle to successful excitation 
energy predictions.
In the present study, we investigate instabilities in variance optimization in order to better 
understand where they come from, whether a pathological variance surface shape is the 
best explanation for these instabilities, and how they may be addressed.

%\textcolor{red}{First, when referring to optimization stability, we mean an optimization's ability to proceed without erroneous parameter changes to the ansatz that yield nonsensical results for the energy or other observables, or change the character of the wave function to that of a different state. Whether an optimization is stable in this sense is a separate issue from whether it has fully converged to the optimal parameter values or if it has encountered a local minimum on the way to the desired global optimum.}
The issue of stability in VMC optimization is multi-faceted, with technical considerations 
about the choice of objective function, the particular optimization algorithm used, the 
level of sampling effort, and the quality of wave function ansatz each playing a role.
Deficiencies in any of these areas could potentially lead to failures of optimization.
For example, if an ansatz is too approximate, it may be that the variance
minimum for the desired state no longer exists, or is so shallow that any method using
statistically uncertain steps will leave the minimum and descend to another.
Even if the minimum exists, an optimization with poor step control may take
undesirable steps that move it out of the desired minimum's basin of convergence.
Both of these issues will show the same end result: an optimization that shows
a decreasing variance, but major changes to the wave function and energy, such as
the large energy rises recently reported. \cite{Cuzzocrea2020}
Whether one or both issues are present can be investigated by introducing increasingly
careful optimization steps through step control algorithms such as trust-radius
schemes.  Of course, no approach to optimization is likely to be successful
in absolutely all circumstances.
However, improvements to optimization algorithms should help limit
optimization failures to cases where the limitations of the ansatz really have
eliminated the desired state's variance minimum.
%relevant question for VMC practitioners is whether these instances of 
%optimization failure pose an 
%insurmountable barrier in practice or whether they can be remedied with certain choices in methodological protocol.

At a broad conceptual level, the objective function being minimized and the particular 
wave function ansatz set the difficulty of a VMC optimization.
Compared to the energy, the variance landscape in parameter space is more difficult to 
resolve at a given level of sampling effort and estimates of its
derivatives may also be more uncertain.\cite{Trail2008a,Trail2008b,Cuzzocrea2020}
While increasing the amount of sampling enables systematic improvement of these 
estimates, the availability of computational resources inevitably places some practical 
upper limit.
The shape of the landscape will also vary with the particular wave function form that 
is chosen.
An isolated minimum of zero variance is guaranteed for any non-degennerate exact eigenstate and some minimum can 
be expected to form when a wave function is systematically built up to this limit, as in 
a Slater determinant expansion.
However, the depth and shape of the minimum are difficult to characterize and the ansatz 
parameters are unlikely to be set to the optimal values initially in practice.
As a result, the particular optimization algorithms chosen to navigate the parameter 
landscape must be highly robust to noisy gradients and capable of handling 
increasing numbers of parameters.
We shall see that different types of algorithms with varying levels of step control 
can be more or less prone to take poor steps when the 
number of parameters is large and statistical noise affects their updates.
Indeed, poor steps due to poor step control appear to be a leading factor behind optimization
instabilities, making more robust step control highly effective.

To further elucidate the issue of optimization stability in VMC, we conduct systematic 
investigations of the LM in the case of the excited state of the cyanine 
dye molecule CN5 that has previously shown instabilities.
We focus on identifying circumstances where variance minimization with the LM fails and what can be 
done to resolve these failures.
Beyond technical choices in operating the LM, we also consider the stability advantages 
offered by a hybrid combination of the LM's blocked variant and AD, particularly 
those rooted in adaptive step control.
As we will see, multiple optimization choices play a role in optimization effectiveness, but the key
determiner of stability appears to be adaptive step control, analogous to using an adaptive trust radius.
In particular, we do not find any indications that variance optimization instabilities arise from pathologies in the 
objective function surface. If this were the case, all optimization 
methods regardless of adaptive step control would eventually show instabilities as well.
As our data below demonstrates, even cases previously reported as unstable become stable 
when sufficiently robust adaptive step control is applied.

\section{Theory}

\subsection{Objective Functions in VMC}

Analysis of optimization stability within VMC begins with consideration of the objective or 
target function being minimized.
A variety of choices have been employed by VMC practitioners, for both ground and excited 
state studies, with the most common choices being the energy or the variance of the energy 
for the system at hand.

\begin{equation}
\label{eqn:ch5_energy}
  E(\vec{\textbf{p}}) = \frac{\int \Psi^2_T(\vec{\textbf{p}})E_L(\vec{\textbf{p}})d\vec{\textbf{R}}}{\int \Psi^2_T(\vec{\textbf{p}})d\vec{\textbf{R}}}
\end{equation}

\begin{equation}
\label{eqn:variance}
\sigma^2_E(\vec{\textbf{p}}) = \frac{\int \Psi^2_T(\vec{\textbf{p}})(E_L(\vec{\textbf{p}}) - E(\vec{\textbf{p}}) )^2 d\vec{\textbf{R}}}{\int \Psi^2_T(\vec{\textbf{p}})d\vec{\textbf{R}}}
\end{equation}

In the above definitions, $E_L = \frac{H\Psi(\textbf{R})}{\Psi(\mathbf{R})}$ is the local 
energy.
Historically, numerical instabilities in early VMC energy minimizations motivated the use of 
variance minimization,\cite{Umrigar1988,Kent1999,Foulkes2001} but later algorithmic 
improvements, particularly the development of the linear method,\cite{Umrigar2007} have 
swung the pendulum back to energy minimization as the main approach for ground state 
studies.
The choice of objective function is less settled for the study of excited states and both energy-based\cite{Cordova2007,Filippi2009,Send2011,Guareschi2013,Guareschi2016,Cuzzocrea2020,Dash2021} 
and variance-based\cite{Umrigar1988,Zhao2016,Shea2017,Flores2019,Garner2020,Otis2020} objective functions have been employed in this context.

\begin{equation}
\label{eqn:ch5_targetAbove}
    \Omega (\Psi) = \frac{\Braket{\Psi | (\omega - H) | \Psi}}{\Braket{\Psi | (\omega - H)^2 | \Psi}}
\end{equation}

\begin{equation}
\label{eqn:ch5_targetClosest}
    W (\Psi) = \Braket{\Psi | (\omega - H)^2 | \Psi}
\end{equation}

\begin{equation}
\label{eqn:ch5_stateAverageE}
    E^{SA} = \sum_I w_I\frac{\Braket{\Psi^I | H | \Psi^I}}{\Braket{\Psi^I  | \Psi^I}}
\end{equation}

The functions $\Omega$ and $W$ both use an input energy parameter $\omega$ to ensure 
targeting of a desired excited state. 
While $\omega$ is held fixed when performing an update to wave function parameters, it must 
ultimately be adjusted to transform $\Omega$ and $W$ into the variance $\sigma^2$ to obtain 
size-consistent results.\cite{Shea2017}
For $\Omega$, the value $\omega$ must eventually be set to the difference of the energy and standard 
deviation, $E - \sigma$, while for $W$, $\omega$ must eventually be set to the energy $E$.
There are multiple strategies for varying $\omega$, either allowing it to float based on the 
values of $E$ and $\sigma$ over the course of a 
single VMC optimization, or performing multiple VMC optimizations at fixed $\omega$ and 
changing the value after each one until self-consistency is achieved.
From the standpoint of optimization stability, allowing the value of $\omega$ to vary is 
one potential source of instability as stochastic uncertainties in the value assigned to 
$\omega$ could possibly lead to the targeting of a different state, and so we investigate this 
possibility below.
We also note that there are various options in the details of varying $\omega$, such as 
allowing it to float only after a period of interpolation from its initial value\cite{Shea2017} or controlling its value through a running average.

The effectiveness of stochastic estimation of different quantities is another factor in 
the stability implications of the choice of objective function.
Sampling from the commonly chosen distribution $\rho (\mathbf{R}) = \frac{\Psi(\mathbf{R})^2}{\int d \mathbf{R} \Psi (\mathbf{R})^2}$ has a zero 
variance property\cite{Assaraf1999} when $\Psi$ is an 
exact eigenstate, but the use of approximate $\Psi$ in practice leads to an infinite 
variance problem for the estimation of the variance itself as well as target function 
derivatives.\cite{Trail2008a,Trail2008b}
While the infinite variance of the variance makes optimization of variance-based target 
functions more difficult, multiples approaches for mitigating the issue have been 
developed, including the use of other importance sampling functions,\cite{Attaccalite2008,Trail2008a,Trail2008b,Robinson2017,Flores2019,Otis2020}
modifications to estimators,\cite{Assaraf1999,Assaraf2000,Assaraf2003}, and 
regularization schemes.\cite{Pathak2020,vanRhijn2022}
We employ an importance sampling function that we have successfully used 
in the minimization of the $\Omega$ objective function in other work.\cite{Otis2020,Otis2021}

\begin{equation}
\label{eqn:ch5_guiding}
    |\Phi|^2 = |\Psi|^2  + c\sum_{j}|\Psi^j|^2 
\end{equation}

In $  |\Phi|^2 $, the coefficient $  c$ weights the sum of squares of wave function parameter derivatives
for the CI parameters and we set it 0.0001 throughout our results.

\subsection{Parameter Optimization Algorithms} \label{algorithms}

A variety of algorithms have been developed for performing parameter optimization in the 
context of VMC.
Some of the most notable ones include the Newton method,\cite{Casalegno2003,Umrigar2005,Sorella2005,Toulouse2008,Cuzzocrea2020} the linear method (LM),\cite{Nightingale2001,Umrigar2007,Toulouse2007,Toulouse2008,Sabzevari2020} stochastic reconfiguration, (SR)\cite{Sorella2007} and accelerated descent (AD) approaches.\cite{Schwarz2017,Sabzevari2018,Luo2019,Mahajan2019,Otis2019}
We focus our discussion on only the aspects of the LM and AD that are most relevant for optimization stability. Further details on
these methods can be found in the literature.\cite{Otis2019,Schwarz2017,Sabzevari2018,Luo2019,Mahajan2019,Nightingale2001,Umrigar2007,Toulouse2007,Toulouse2008,Zhao2017}

The LM relies on a first order Taylor expansion of the trial wave function and in the 
case of minimizing $\Omega$  leads to a generalized eigenvalue problem

\begin{equation}
\label{eqn:ch5_lmEigen}
    (\omega - \mathbf{H}) \hspace{0.6mm} \mathbf{c}
    = \lambda \hspace{0.6mm} (\omega - \mathbf{H})^2 \hspace{0.6mm} \mathbf{c}
\end{equation}

with matrix elements of the form

 \begin{equation}
 \label{eqn:ch5_lmLHS}
     \Braket{\Psi_i | \omega - H | \Psi_j}
 \end{equation}
 and
 \begin{equation}
 \label{eqn:ch5_lmRHS}
     \Braket{\Psi_i | (\omega - H)^2 | \Psi_j}.
 \end{equation}
where $\Psi_i$ and $\Psi_j$ are wave function derivatives with respect to variational 
parameters.
The LM requires stochastic estimates of these matrix elements to compute an update 
$\mathbf{c}$ to the wave function parameters.
The size of the LM matrices will grow as the square of the number of parameters.
For a fixed number of statistical samples, increasing the number of parameters can
increase the uncertainty in the LM's parameter update steps, which in turn increases the 
risk of optimization failure.

\begin{figure}[H]

\includegraphics[width=\textwidth]{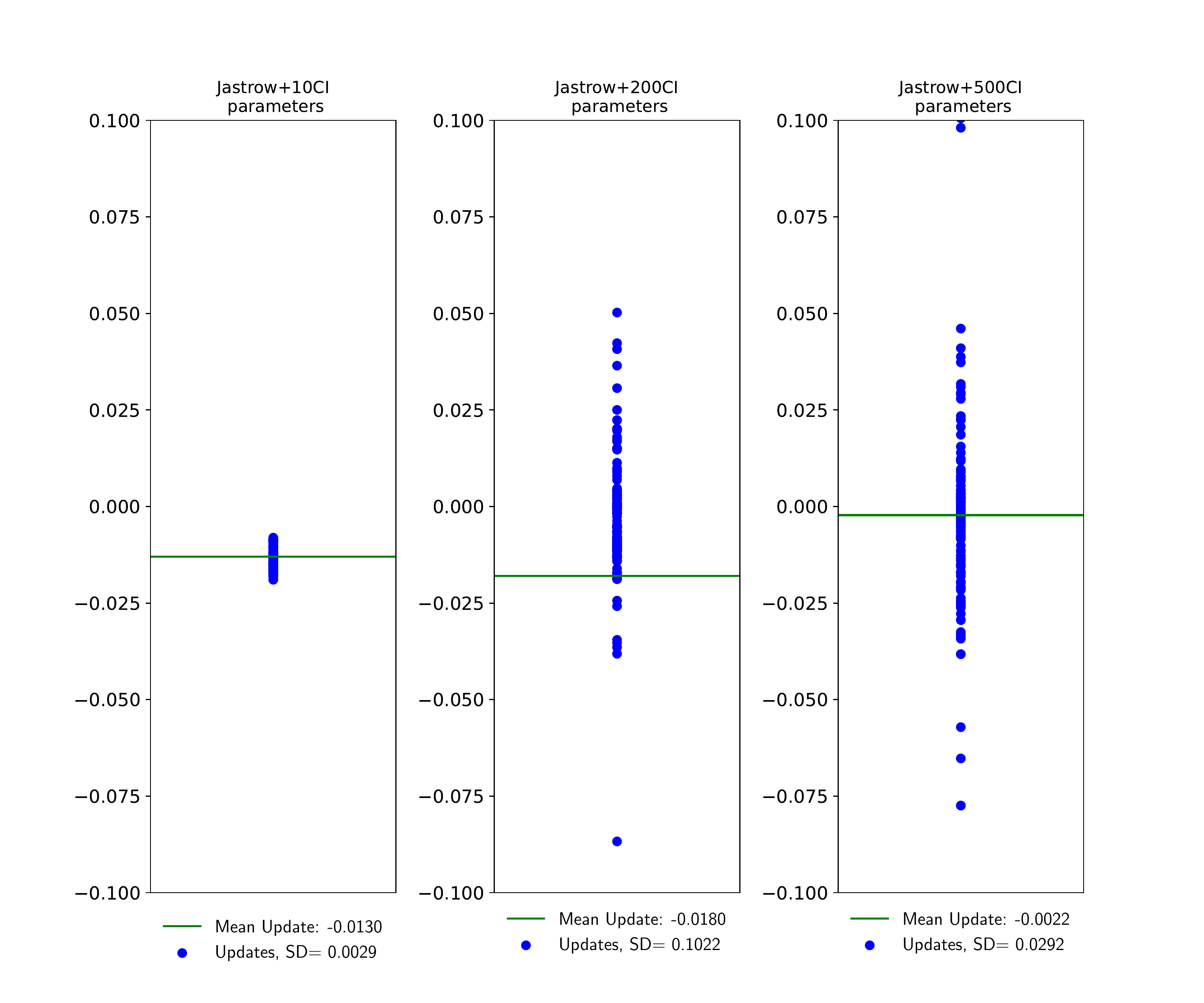}
\caption{Proposed LM updates to a Jastrow spline parameter for varying total numbers of parameters.   Each panel was produced with 200 proposed updates from independent LM runs on an excited state wave
function in CN5. The same scale has been used in each panel, which leaves off some larger outliers in the 200 and 500 determinant cases. From left to right, the total numbers of wave function parameters for the different cases are 150, 340, and 640. 100,000 samples were used to generate the LM matrices for each update.}
\label{fig:lm_no_shift_update}

\end{figure}

An example illustrating this issue is given in Figure \ref{fig:lm_no_shift_update}, which 
shows the spreads of proposed LM updates on the same Jastrow parameter when increasing 
total numbers of parameters are included at a fixed statistical sample size.
As more CI parameters are added, the statistical spread of the updates to this Jastrow 
parameter increases by an order of magnitude.
What is going on here is that the LM's eigenproblem effectively amplifies and couples the 
noise in the updates for individual parameters.
If this were not the case, the spread in the Jastrow update would remain constant regardless 
of how many parameters were included.
Given that the degree of nonlinearity present in a diagonalization, reflected in the degree of the characteristic polynomial, grows with matrix 
dimension, it is not entirely surprising that diagonalizations of larger and larger 
sets of statistically uncertain matrix elements lead to increased uncertainty in 
the proposed parameter updates.
For larger numbers of parameters, naive acceptance of these increasingly uncertain LM updates
could lead to instabilities in optimization, which strongly suggests that step control 
methods that work to reduce this uncertainty and that adapt their behavior based on the 
degree of uncertainty will be valuable.

\begin{figure}[H]

\includegraphics[width=\textwidth]{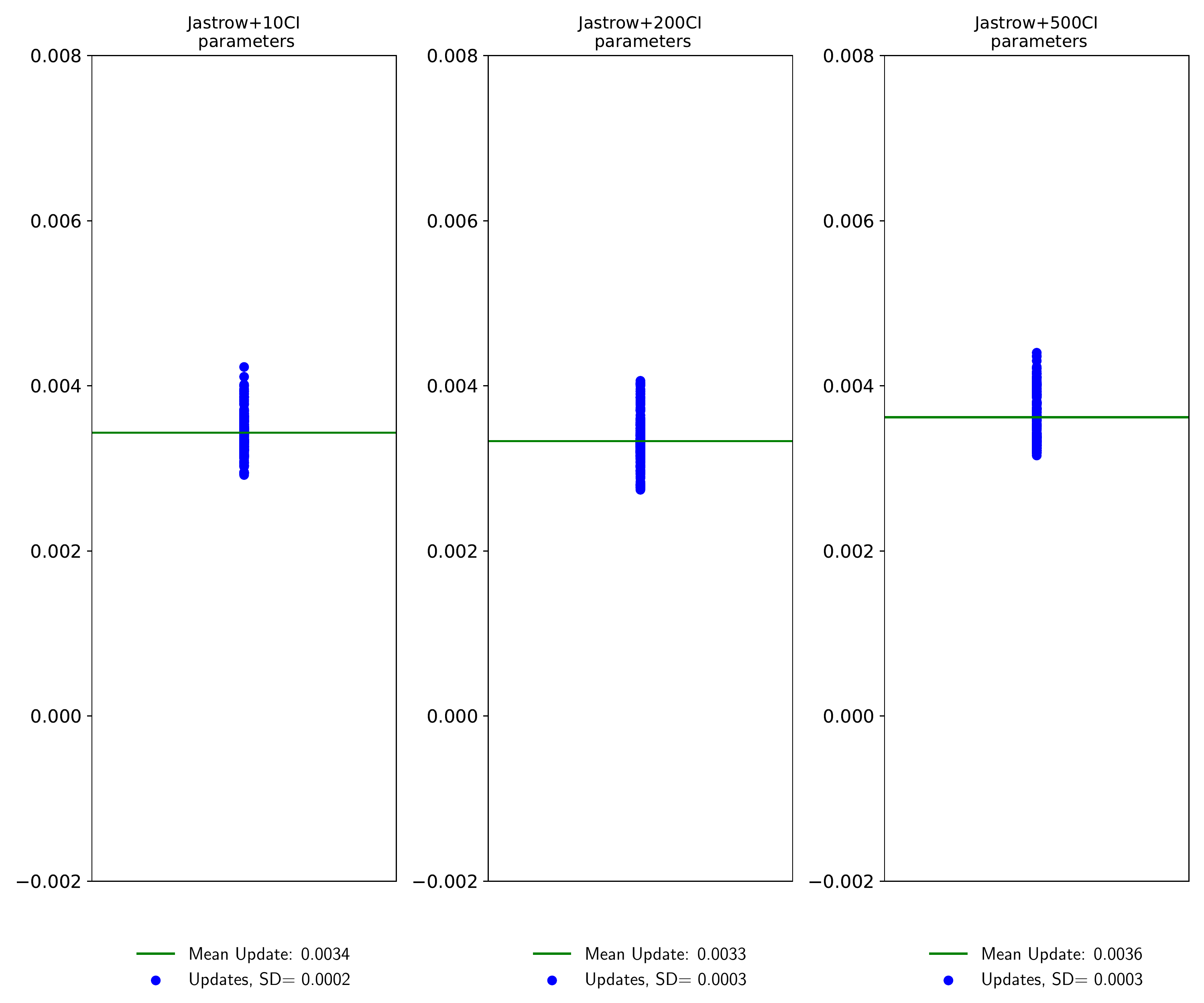}
\caption{Proposed LM updates to a Jastrow spline parameter with a diagonal shift of 100 for varying total numbers of parameters. Each panel shows 200 proposed updates from independent LM runs with an excited state wave
function in CN5. 100,000 samples were used to generate the LM matrices for each update.}
\label{fig:lm_shift_updates}

\end{figure}

A number of technical modifications to the basic LM algorithm can reduce, though not 
eliminate, this problem.
Shift values can be added to the LM Hamiltonian matrix to prevent overly large parameter 
changes, similar to the use of trust radius schemes in Newton-Raphson.\cite{Kim2018,Otis2019}
Two different types of shifts, defined in equations \ref{eqn:lmShift} through \ref{eqn:tMat}, 
can help address distinct problems in the LM optimization.
The diagonal shift $c_I$ helps stabilize the LM by effectively adding an energy penalty to 
parameter changes to the current wave function.\cite{Umrigar2007}
While this shift reduces the effective step size of the LM, it also changes the direction of 
the LM updates by rotating them to the steepest descent direction in the limit of an infinite
shift.\cite{Toulouse2007}
Figure \ref{fig:lm_shift_updates} demonstrates the effect of a large diagonal shift in 
controlling the noise in parameter updates.
The spread of proposed updates for the Jastrow parameter is very small and also remains 
constant across the different total numbers of parameters, which is the behavior 
expected in steepest descent.
It is important to emphasize that the step damping parameter used in the previous 
CN5 study\cite{Cuzzocrea2020} does not have this effect.
It only scales the step without modifying the step's direction, meaning that 
when using it without any other step control method, the phenomenon of step 
uncertainty increasing with parameter number will still be present.
Without the ability to curb the possibility of poor step directions, adjusting the size of this scaling factor would merely change the number of iterations over which instabilities occur, as has indeed been 
observed.\cite{Cuzzocrea2020}

While the diagonal shift prevents excessive changes in parameters with large derivatives, 
assigning it a large enough value for this task may turn off the optimization of parameters 
with smaller, but still significant gradients.
To avoid this scenario, a $c_S\mathbf{B}$ term (defined in terms of matrices $\mathbf{T}$ and $\mathbf{Q}$ in Equations \ref{eqn:bMat} through \ref{eqn:tMat})  can be added to $\mathbf{H}$ to help retain flexibility in other parameter 
directions with smaller derivatives.
The overlap shift $c_S$ imposes an energy penalty based on the norm of parameter 
directions orthogonal to the current wave function so that steps in the directions of 
small derivatives are penalized less heavily and meaningful changes in them can still occur.
In practice, the matrix $\mathbf{B}$ does not need to be constructed explicitly, as the 
LM generalized eigenvalue problem can be solved by Krylov iteration.\cite{Kim2018,Sabzevari2020}

\begin{equation}
\label{eqn:lmShift}
    \mathbf{H} \xrightarrow[]{} \mathbf{H} + c_I \mathbf{A} + c_S \mathbf{B}
\end{equation}

\begin{equation}
    A_{ij} = \delta_{ij}(1-\delta_{i0})
\end{equation}

\begin{equation}
\label{eqn:bMat}
\mathbf{B} = (\mathbf{Q^T})^{-1} \mathbf{T} \mathbf{Q}^{-1}
\end{equation}

\begin{equation}
\label{eqn:qMat}
    Q_{ij} = \delta_{ij} -\delta_{i0}(1-\delta_{j0})S_{0j}
\end{equation}

\begin{equation}
\label{eqn:tMat}
    T_{ij} = (1-\delta_{i0}\delta_{j0})[\mathbf{Q^T}\mathbf{S}\mathbf{Q}]_{ij}
\end{equation}

In our own implementation of the linear method, the use of shifts is enhanced in 
combination with an adaptive scheme\cite{Umrigar2007,Toulouse2007,Toulouse2008} where three sets of shift values are used to determine 
parameter updates, and a correlated sampling procedure either selects the candidate update 
expected to reduce the objective function the most, or rejects all the possible updates and 
raises the shifts on the next LM iteration for more cautious steps.
This allows the step control to account in at least a limited way for the amount of 
statistical uncertainty present in a given stage of the optimization.
Despite this ability to reject poor steps and the stability benefits it offers, we will show in our results that this approach to the LM can still 
fail to lower the objective function effectively in some circumstances.

A more substantial modification to the LM is to divide the parameters into blocks and 
solve a set of smaller eigenvalue problems to produce the parameter update.\cite{Zhao2017}
The blocked LM algorithm consists of two phases.
First, an LM-style diagonalization is performed for each of the $N_b$ blocks of parameters 
and the $N_k$ lowest eigenvectors are retained from each block.
In the second phase, the full set of $N_b N_k$ eigenvectors plus $N_o$ other parameter 
directions (either from previous blocked LM iterations or descent, in the case of the
hybrid method) form the space for a final LM diagonalization that gives the overall update.
This approach provides the blocked LM with a lower memory footprint compared to the 
standard LM, but reduces the flexibility of the directions it can explore in 
parameter space.
By working with smaller matrices, however, the blocked LM reduces the degree of nonlinear coupling in its equations and thus directly addresses the issue shown in Figure \ref{fig:lm_no_shift_update}.
It therefore has the potential to provide more reliable update steps.
However, in the limit of more variational parameters with insufficient sampling effort, the
blocked LM will share the same vulnerabilities as the standard LM.

AD methods are a class of pseudo-first-order optimization algorithms that use 
only the gradient of the target function, retaining some memory of gradients from previous 
iterations in order to converge more quickly than steepest descent.
Many different flavors of AD have been widely used in the machine learning community and 
a number of them have been applied in the context of VMC.\cite{Schwarz2017,Sabzevari2018,Luo2019,Mahajan2019,Otis2019}
Below, we give the defining equations for a combination of RMSprop and Nesterov momentum as 
used in a number of VMC studies\cite{Schwarz2017,Otis2019,Otis2020,Otis2021} and which we
use here in the hybrid method.

 \begin{equation}
\label{eqn:ch5_rmspropMomentum}
    p_i^{k+1} = (1-\gamma_k e^{-(\frac{1}{d})(k-1)})q_i^{k+1} - \gamma_k e^{-(\frac{1}{d})(k-1)} q_i^k
\end{equation}
\begin{equation}
\label{eqn:ch5_rmspropUpdate}
   q_i^{k+1} = p_i^k - \tau_k \frac{\partial \Omega (\mathbf{p})}{\partial p_i}
\end{equation}
\begin{equation}
\label{eqn:ch5_rmspropRecur}
\lambda_0=0 \hspace{7mm}
\lambda_k = \frac{1}{2} + \frac{1}{2}\sqrt{1+4\lambda_{k-1}^2} \hspace{7mm}
\gamma_k = \frac{1-\lambda_k}{\lambda_{k+1}}
\end{equation}
\begin{equation}
\label{eqn:ch5_rmspropStep}
    \tau_k = \frac{\eta}{\sqrt{E[(\frac{\partial\Omega}{\partial p_i})^2]^{(k)}} + \epsilon}
\end{equation}
\begin{equation}
\label{eqn:ch5_rmspropAvg}
    E[(\partial \Omega)^2]^{(k)} = \rho E\left[\left(\frac{\partial \Omega}{\partial p_i}\right)^2\right]^{(k-1)} + (1-\rho)\left(\frac{\partial\Omega}{\partial p_i}\right)^2
\end{equation}
The recurrence relations defined in equations \ref{eqn:ch5_rmspropMomentum} through 
\ref{eqn:ch5_rmspropRecur} define the momentum in this algorithm.
Obtaining the updated parameter $p_i^{k+1}$ depends on the values of the target function 
gradient $\frac{\partial \Omega (\mathbf{p})}{\partial p_i}$ from both the present and previous iterations, which accelerates the convergence compared to steepest descent.
Equations \ref{eqn:ch5_rmspropStep} and \ref{eqn:ch5_rmspropAvg} define the RMSprop adaptive 
modification of the step size $\tau_k$, which is parameter-specific.

From a stability perspective, AD approaches offer a number of advantages compared to the LM.
The relevant equations are far more linear than the LM's generalized eigenvalue problem, which 
reduces the dangers of step bias and uncertainty from nonlinear combinations of stochastic 
quantities.
In practice, VMC optimization with AD can be conducted with a significantly lower per-iteration
sampling effort than the LM without creating instabilities.
While an individual AD step may be poor and raise the objective function, modest step sizes 
ensure that the change to the wave function is small, and typical AD optimizations use many 
hundreds or thousands of iterations to successfully minimize the objective function.
However, these stabilizing features of AD also leave it slower to converge to the minimum than the LM.\cite{Otis2019}
This weakness has motivated the development of a hybrid approach that alternates between 
sections of AD and blocked LM optimization, in order to allow the blocked LM to more swiftly 
move parameters to their optimal values.\cite{Otis2019,Otis2020,Otis2021}

\subsection{Wave Functions}

Another key aspect of VMC stability is the choice of wave function, which shapes the target 
function landscape and thus the difficulty of the optimization problem.
With the exact wave function and an infinite sample size, variance minima will exist for all 
the non-degenerate Hamiltonian eigenstates.\cite{Umrigar1988}
For approximate ansatzes and finite sampling effort, the minima may become shallow enough 
that some optimizers fail to target and converge to them or they may not exist at all.
This interplay between wave function quality and optimizer effectiveness
can lead to varying stability outcomes for different ansatz choices as has previously been observed.\cite{Cuzzocrea2020}
Therefore, as a practical matter, being able to systematically improve the wave function and 
the amount of sampling may make the difference between success and failure for a given 
optimization algorithm.
We have provided a recent example of this by applying the LM to simple wave functions in CN5,
where adding a 3-body Jastrow factor and increasing sampling eliminated 
instabilities.\cite{Otis2020}

In this work, we consider more complex wave functions in CN5, all of the Multi-Slater 
Jastrow (MSJ) form shown below.
The Slater determinant expansion $\psi_{MS}$ can be obtained from various active space 
calculations of different sizes, with larger spaces potentially solved by sCI, producing 
longer and more accurate expansions.

\begin{equation}
\label{eqn:ch5_psi}
    \Psi = \psi_{MS} \psi_J 
\end{equation}
\begin{equation}
\label{eqn:ch5_psiMS}
    \psi_{MS} = \sum_{i=0}^{N_D} c_i D_i
\end{equation}
\begin{equation}
\label{eqn:ch5_psiJ}
    \psi_J = \exp{\sum_i \sum_j \chi_{ij}(|r_i - R_j|) + \sum_k \sum_{l>k} u_{kl} (|r_k - r_l|)}
\end{equation}

Throughout our results, we employ and optimize one- and two-body Jastrow factors, which are 
constructed with splines for the functions $\chi_k$ and $u_{kl}$.\cite{Kim2018}
For our stability analysis, we also optimize the orbitals of our Slater expansions, 
benefiting from recent methodological improvements with the table method.\cite{Filippi2016,Assaraf2017}
However, as we show in our results, orbital optimization can still be challenging for the LM, 
and we also consider VMC optimization with the orbitals left at shapes obtained from a 
recent state-specific CASSCF approach.\cite{Tran2019,Tran2020,Hanscam2021}
This approach of combining state-specific quantum chemistry with VMC has recently been shown
to provide accurate excitation energies across a range of different types of excited 
states\cite{Otis2021} and we provide another example of this here for the excited state in 
CN5.
However, the key thrust of the present study is optimization stability, and so in many of our results we will be optimizing orbitals within VMC to test stability in that context.

\section{Results}

\subsection{Computational Details}

We perform all VMC optimizations in a development version of QMCPACK.\cite{Kim2018,Kent2020}
Our molecular geometry is identical to one recently used by Filippi and 
coworkers\cite{Cuzzocrea2020,Boulanger2014} and we repeat the coordinates in the appendix 
for convenience.
Basis set and pseudopotential choices are specified below for the particular wave function 
cases we consider.
For generating our wave functions, we have performed CASSCF calculations in Molpro\cite{MOLPRO} and PySCF,\cite{Sun2018} 
as well as sCI calculations in Dice.\cite{Holmes2016,Sharma2017}
Our one- and two-body Jastrow factor splines each consist of 10 coefficients defining the 
function within a cutoff distance of 10 bohr.

\subsection{Excitation Energies from Variance Minimization}

We begin by simply checking the ability of variance-based excited state VMC to obtain 
accurate excitation energies with the hybrid method.
For this test, we follow the same methodology that has recently been applied on a broader set of 
excited states.\cite{Otis2021}
We use a cc-pVTZ basis set with pseudopotentials\cite{Bennett2017} and obtained state-specific 
CASSCF orbitals in a (6e,5o) space.
We then generate Slater determinants from heatbath CI in a (28e,40o) space of these orbitals.
The orbitals are left at their state-specific CASSCF shapes and we optimized only Jastrow parameters and 
CI coefficients in VMC with the hybrid method, using a recent parameter selection 
scheme\cite{Otis2021} to optimize only parameters with significant derivatives within the 
blocked LM.
For the blocked LM portions, we divided those parameters into 5 blocks and retained 30 
parameter directions per block along with 5 directions from AD for the second phase of the 
blocked LM algorithm.
We alternate between 100 iterations of AD and 3 iterations of the blocked LM and refer to the combination of the two sections as a macro-iteration of the hybrid method.
No stability issues were observed in these optimizations.
To ensure balance when taking energy
differences, we employ explicit variance matching.
We perform this overall procedure for two pairs of
wave functions:  the first has 100 determinants for the
ground state and 1000 for the excited state, and the
second has 300 for ground and 1500 for excited.

\begin{figure}[H]

\includegraphics[width=\textwidth]{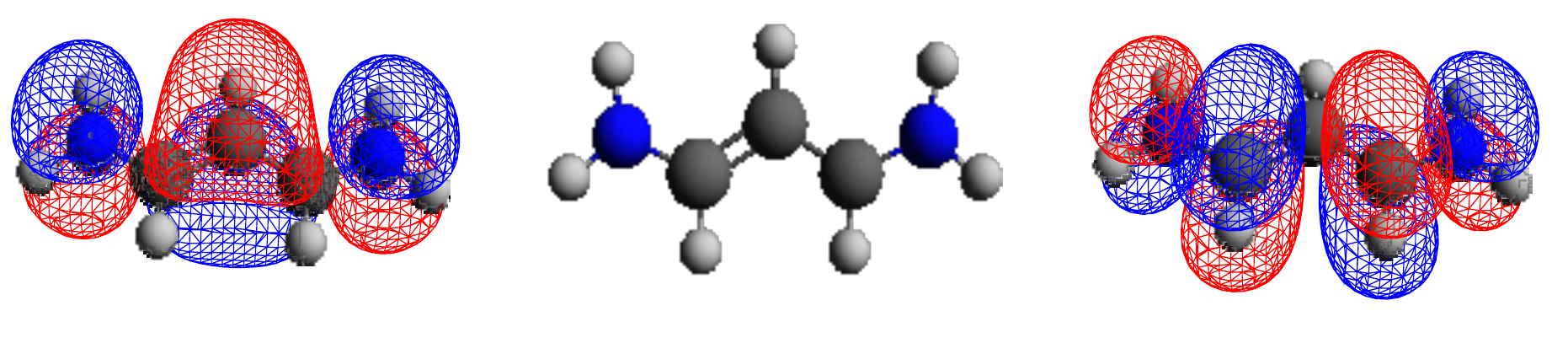}
\caption{Structure (center) for CN5: \ce{C_3H_3 (NH_2)_2^+}. Hole (left) and particle (right) orbitals for the CN5 excited state.}
\label{fig:cn5_orbs}

\end{figure}

\begin{figure}[H]

\includegraphics[width=\textwidth]{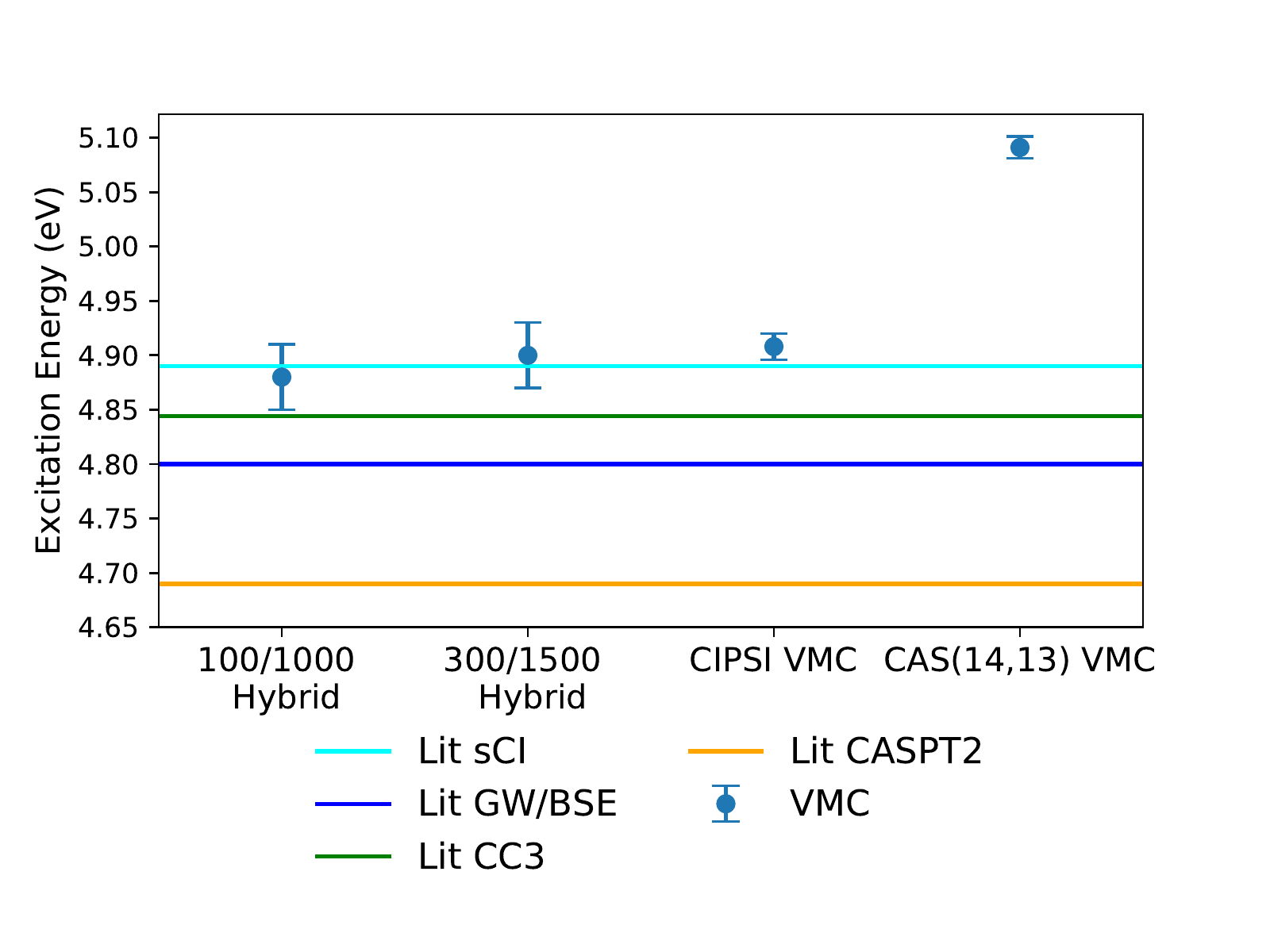}
\caption{Excitation energy in CN5 using the hybrid method to minimize the $\Omega$ target function. Reference values for  and CAS(14,13) VMC,\cite{Cuzzocrea2020} CC3,\cite{Cuzzocrea2020} CASPT2,\cite{Boulanger2014,Send2011} GW/BSE,\cite{Boulanger2014} and sCI\cite{Garniron2018} are taken from the literature. For comparison, the basis sets used in the literature results 
were aug-cc-pVTZ for CC3 and GW/BSE, aug-cc-pVDZ for sCI, ANO-L-VTZP for CASPT2, and a double-$\zeta$ basis set minimally augmented with $s$ and $p$ diffuse functions on heavy atoms for SA-VMC and CIPSI VMC.}
\label{fig:cn5_excitation_energies}

\end{figure}

\begin{table}[H]

\caption{Excitation energies for CN5 for VMC and literature results.\cite{Cuzzocrea2020,Boulanger2014,Send2011,Garniron2018} Stochastic uncertainties on the last digit are given in parentheses. The two hybrid method VMC results relied on variance-based minimizations while the two literature VMC values were obtained from energy minimizations using wave functions from a (14e, 13o) CAS and balanced CIPSI expansions.\cite{Cuzzocrea2020} The 100/1000 hybrid method optimizations obtained variances of 0.667(1) a.u. and 0.666(1) a.u. respectively while the 300/1500 optimizations both obtained 0.657(1) a.u.}
\begin{tabular}{lS}
Method & \multicolumn{1}{l}{Excitation Energy (eV)}   \\ \hline

Hybrid 100/1000 VMC    & 4.88(3)        \\
Hybrid 300/1500 VMC    & 4.90(3)        \\
CIPSI VMC  & 4.91(1) \\
CAS(14,13) VMC  & 5.09(1) \\
CASPT2    & 4.69        \\
GW/BSE & 4.80   \\
CC3    & 4.84       \\
sCI  & 4.89            \\

\end{tabular}

\label{tab:cn5_energies_data}
\end{table}

As shown in Figure \ref{fig:cn5_excitation_energies} and Table \ref{tab:cn5_energies_data},
we obtain a highly accurate excitation energy for this state in CN5 for our variance-based 
methodology.
Our result is with 0.1 eV of literature results from sCI\cite{Garniron2018} and 
CC3\cite{Cuzzocrea2020} and statistically indistinguishable from a VMC result from Filippi and 
coworkers based on  energy minimization.\cite{Cuzzocrea2020}
Thus, we have a practical demonstration of variance based minimization with 
the hybrid method achieving an accurate excitation energy without stability concerns for this particular state when relying on state-specific quantum chemistry to provide orbital shapes.
We now turn to more rigorous stability testing, where VMC orbital optimization plays a 
large role.

\subsection{Stability Tests}

To gain more insight into cases where stability issues might arise, we consider a series 
of excited state optimizations with the LM and the hybrid method.
Our focus is now mainly on whether these optimizations progress stably rather than whether they
fully reach the minimum or the precise accuracy of the excitation energies that might be 
obtained from them.
The differences in technical optimization choices between the stable and unstable outcomes 
we observe may assist the work of other VMC practitioners.

To construct an ansatz for this analysis, we consider a (6e,10o) active space in a cc-pVDZ 
basis set with BFD pseudopotentials.\cite{Burkatzki2007}
This is the same basis, pseudopotential and active space for which stability issues were recently 
observed.\cite{Cuzzocrea2020}
We use a 0.001 coefficient cutoff for a 4-state-average CASSCF to obtain a CI expansion of 1892
determinants for the lowest excited state of CN5 and add one- and two-body Jastrow factors.
The lowest excited state is a ground state within its symmetry and could have been described 
with a single state CASSCF, but here we allow determinants of other 
symmetries (which had coefficient above 0.001 for other states in the state-average but 
have zero coefficients for the lowest excited state) within our ansatz.
The presence of these unnecessary determinants increases the difficulty of the optimization 
due to the effect in Figure \ref{fig:lm_no_shift_update}
and the use of state-averaging may enhance the amount of orbital optimization being asked of 
VMC.
We note that both these aspects of the wave function are different from the (6e,10o) CAS ansatz studied by Filippi and coworkers, which included only determinants of the correct symmetry without the use of state-averaging.\cite{Cuzzocrea2020}
While this ansatz is not as closely tailored for the excited state as quantum chemistry 
might allow, it remains a reasonable initial guess that still requires nontrivial 
optimization at the VMC level.

In order to assess stability issues that may appear gradually, we run our optimizations for 
a considerably larger number of iterations than we otherwise would for typical excitation 
energy predictions, aiming for many hundreds of LM iterations.
We also provide the LM (and its blocked variant within the hybrid method) with only 100,000 samples
per iteration in order to make long optimizations more practical and somewhat favor the 
potential emergence of stability issues due to poorer stochastic estimates.
For comparison, we note that for a molecule as large as CN5, our typical practices would 
provide a million samples per LM iteration, potentially more in cases with many ansatz 
parameters, and only optimize for somewhere in the range of 50 to 100 iterations.
For the hybrid method, we use 30,000 samples per AD iteration, which is a typical amount, and
use 17 macro-iterations (each containing 100 AD steps and 3 blocked LM steps), which is 
significantly longer than our usual range of 5 to 10.

We begin with a fixed value of $\omega =-41.475$ and minimize the objective function $\Omega$ 
in a staged approach, first optimizing the less difficult Jastrow 
and CI parameters and only turning on orbital optimization at the next stage.
We choose to optimize all determinants from CASSCF alongside the initially unoptimized 
Jastrow for a total of 2030 variational parameters in our first stage of optimization.
We mention that in practical calculations, this stage could be performed even more cautiously, 
such as by optimizing the Jastrow with only the 100 most important determinants and 
then adding the remainder in sets of a few hundred with zero initial coefficients, but optimizing 
them all at once from the start provides us with a more stringent stability test.

Three optimizer choices are considered: LM with fixed shifts, LM with adaptive shifts, and 
the hybrid method.
These various options' approaches to the issue of step size control are a key way of 
distinguishing them.
In both varieties of the LM as well as the blocked LM steps within the hybrid method, a 
correlated sampling procedure is performed to allow the possibility of rejecting 
a proposed parameter update.
However, the fixed shift version of the LM has no means of constraining its step size 
beyond what the chosen shift values($c_I = 0.1$ and $c_S = 1$) provide while the 
adaptive version can move from those initial values and thus alter both the size and 
direction of the LM updates, as explained above in Section \ref{algorithms}.
The hybrid method incorporates the benefits of the adaptive shift scheme in its use of the 
blocked LM and the step sizes employed during its AD sections dynamically adjust with the 
objective function gradients according to RMSprop.
Another detail is that the hybrid method resets the shifts after each macro-iteration for the next set of blocked LM iterations.
Throughout the hybrid method optimizations in this section, we divide parameters into 5 blocks 
and employ 30 parameter directions per block along with 5 directions from AD in the 
second phase of all blocked LM iterations.

\begin{figure}[H]

\includegraphics[width=\textwidth]{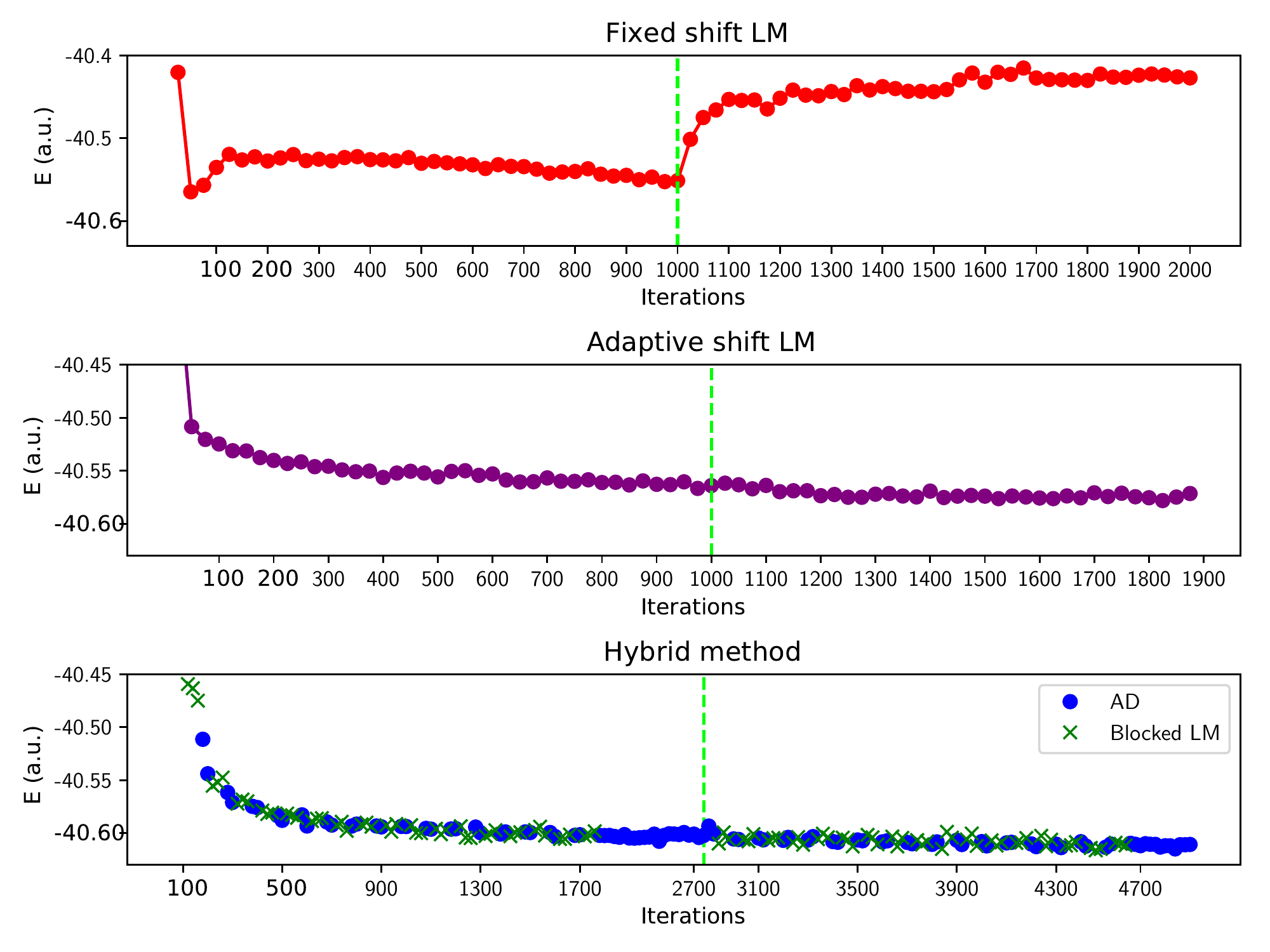} 
\caption{Excited state energy for fixed shift LM, adaptive shift LM, and hybrid method. Starting point was unoptimized Jastrow and CI coefficients from SA-CASSCF. Green dashed line 
marks the point at which orbital optimization is turned on.}
\label{fig:all_methods_energy}

\end{figure}

\begin{figure}[H]

\includegraphics[width=\textwidth]{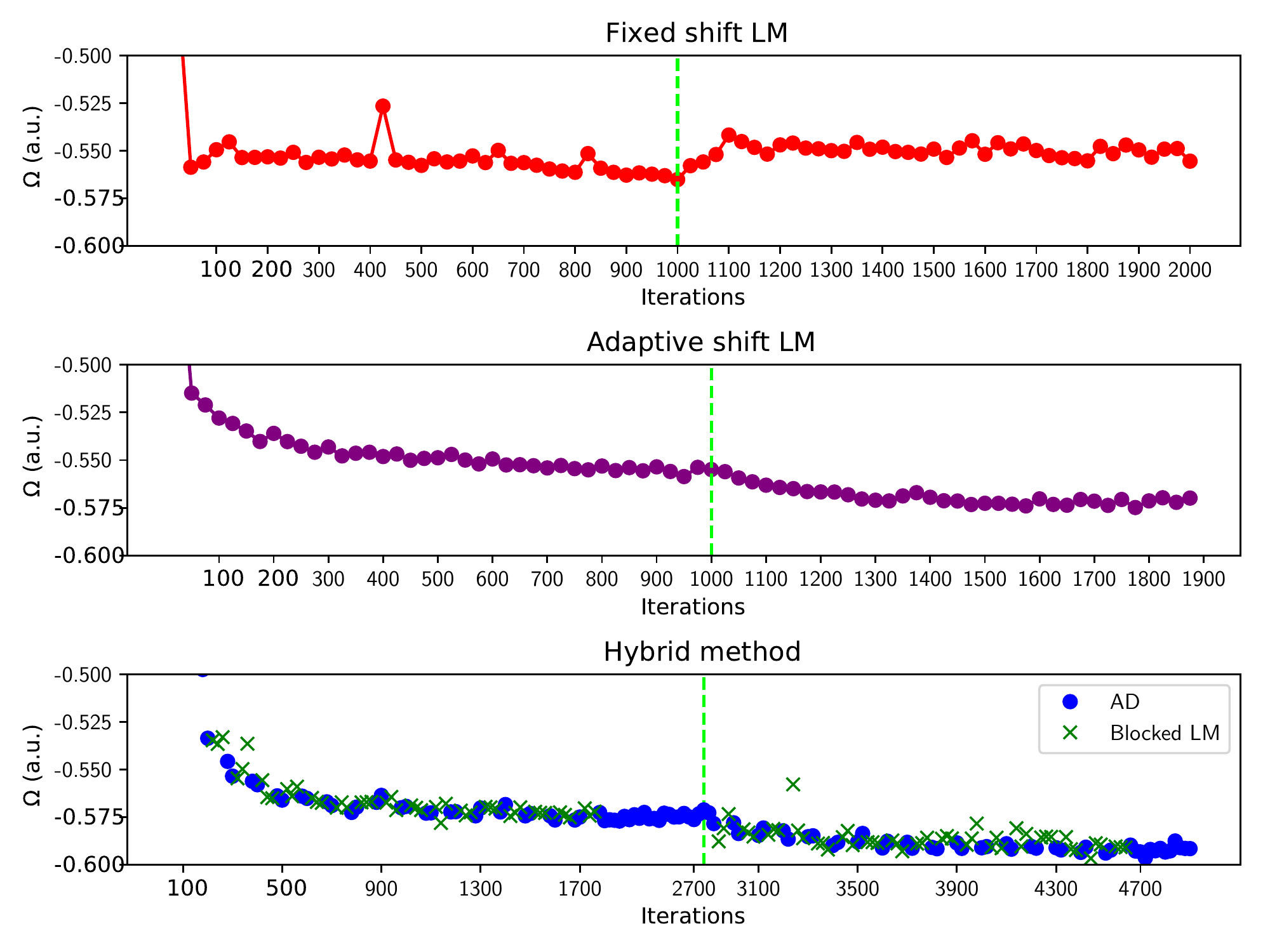} 
\caption{Objective function $\Omega$ for fixed shift LM, adaptive shift LM, and hybrid method. Starting point was unoptimized Jastrow and CI coefficients from SA-CASSCF.Green dashed line 
marks the point at which orbital optimization is turned on}
\label{fig:all_methods_target}

\end{figure}

\begin{figure}[H]

\includegraphics[width=\textwidth]{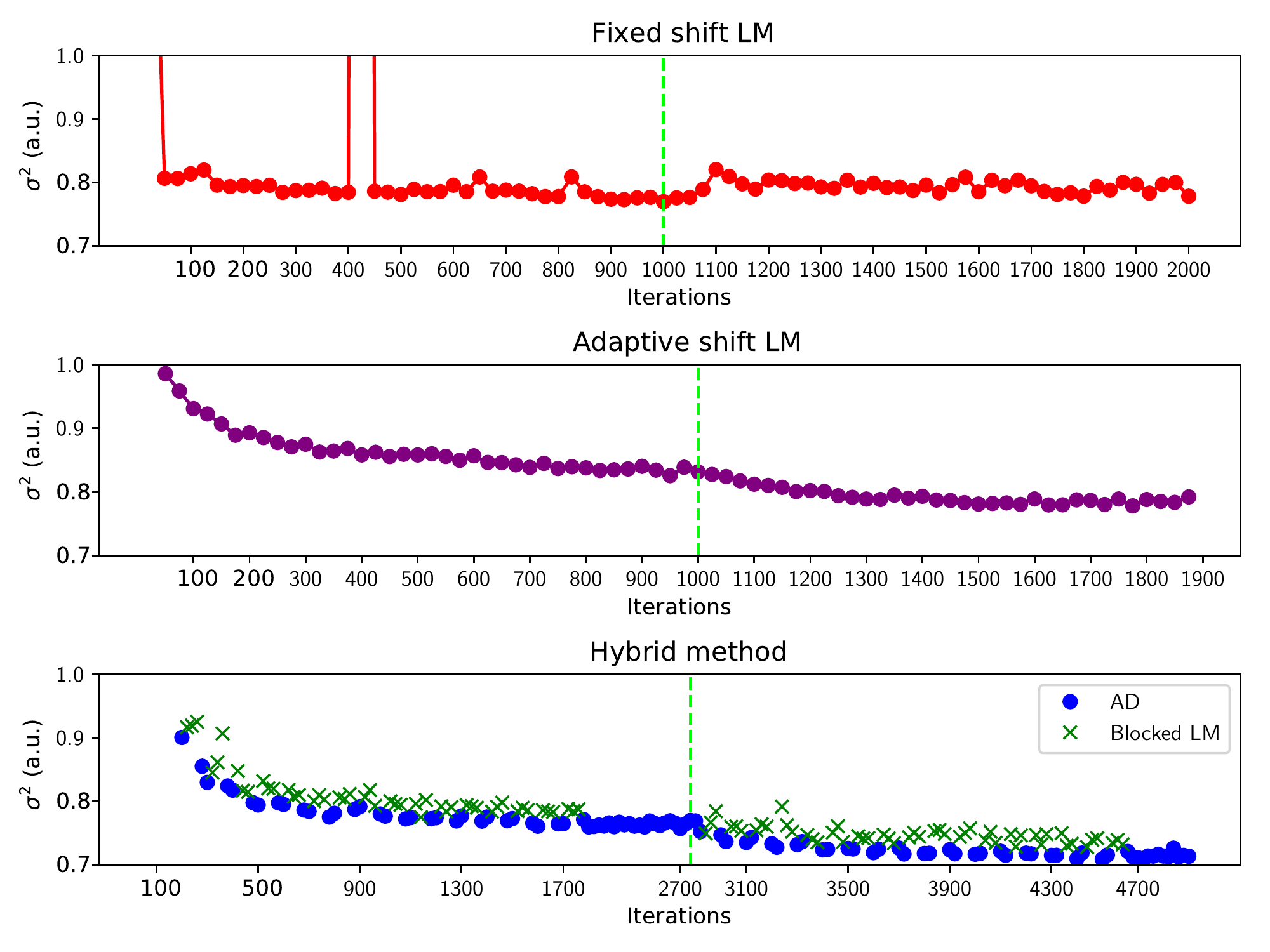} 
\caption{Excited state variance for fixed shift LM, adaptive shift LM, and hybrid method. Starting point was unoptimized Jastrow and CI coefficients from SA-CASSCF. Green dashed line 
marks the point at which orbital optimization is turned on. 
The separation between AD and blocked LM points is an artifact of autocorrelation.}
\label{fig:all_methods_variance}

\end{figure}

Turning to the results, we consider the optimizations shown in Figures \ref{fig:all_methods_energy} through \ref{fig:all_methods_variance}.
Points on the LM plots represent the average of 25 LM iterations.
On the hybrid method plots, dots are averages of 50 AD iterations and crosses are 
individual blocked LM iterations.
The vertical scales have been chosen to make the varying stability behavior of the 
optimization algorithms more visible, which leaves off some points corresponding to early 
iterations.
The vertical dashed line marks the end of the first stage of optimizing only Jastrow 
and CI parameters and the start of including orbital optimization.
At the onset of orbital optimization, the initial values of the shifts were restored to $c_I = 0.1$ and $c_S = 1$ for the 
adaptive shift LM and the hybrid method, and the initial step sizes used in AD are 
changed.
The initial hybrid method optimization of CI and Jastrow parameters used step sizes of 
0.01 and 0.1 respectively.
Once orbital optimization was turned on, these were reduced to 0.005 and orbital parameters 
were given a step size of 0.0001.

Focusing on the left-hand Jastrow and CI side of the three figures, we see that all approaches
successfully lower the target function with the energy and variance also decreasing in 
tandem.
The fixed shift LM case in the upper panel of the three figures provides some reason for 
concern with a sustained rise in the target function and energy within the first 
100 iterations, but the optimization later recovers.
The adaptive shift LM and the hybrid method provide smoother optimization, but for 
this case with only the easier Jastrow and CI parameters, their refinements for 
step size control may be less necessary.

We then turn on orbital optimization and continue to optimize the other parameters starting 
from each method's Jastrow and CI result.
We continue to keep $\omega$ fixed.
With orbital optimization, our ansatz now has a total of 6163 variational parameters.
We now see in the second half of Figures \ref{fig:all_methods_energy} through \ref{fig:all_methods_variance} that the three methods differ more 
dramatically.
The fixed shift LM shows a substantial rise in energy of roughly 80 milliHartree over the 
course of hundreds of iterations.
This failure of optimization is clearly signaled by a rise in the objective function in 
the first hundred iterations of orbital optimization.
Both the energy and objective function remain elevated during hundreds of later 
iterations, indicating that the LM is unable to recover in this case.
While the fixed shift LM in principle possesses the ability in the correlated sampling 
phase to detect and reject any step that would raise the objective function, this 
safeguard is clearly inadequate in this case.
With poor estimates of the objective function at the current and proposed sets of ansatz 
parameters, the LM may mistakenly estimate that a step will lower the objective function only 
to discover otherwise once it is at the new point in parameter space and takes new 
samples.
This danger is exacerbated if proposed parameter
steps can remain incautiously large and 
uncertain, which fixing the LM shifts permits.
Although our fixed-shift LM
and the previously employed \cite{Cuzzocrea2020}
fixed-damping Newton method differ in their particulars
and in the details of how the instabilities manifest,
they both fit into a broad pattern: methods that rely on
a statistical estimate of a second-derivative matrix
without also employing adaptive step control display
unstable behavior during excited state variance optimization.
If anything, the fixed-shift LM is even more
unstable, as its steps are poor to the point that they raise
the objective function during the orbital
optimization stage.

\begin{figure}[H]

\includegraphics[width=\textwidth]{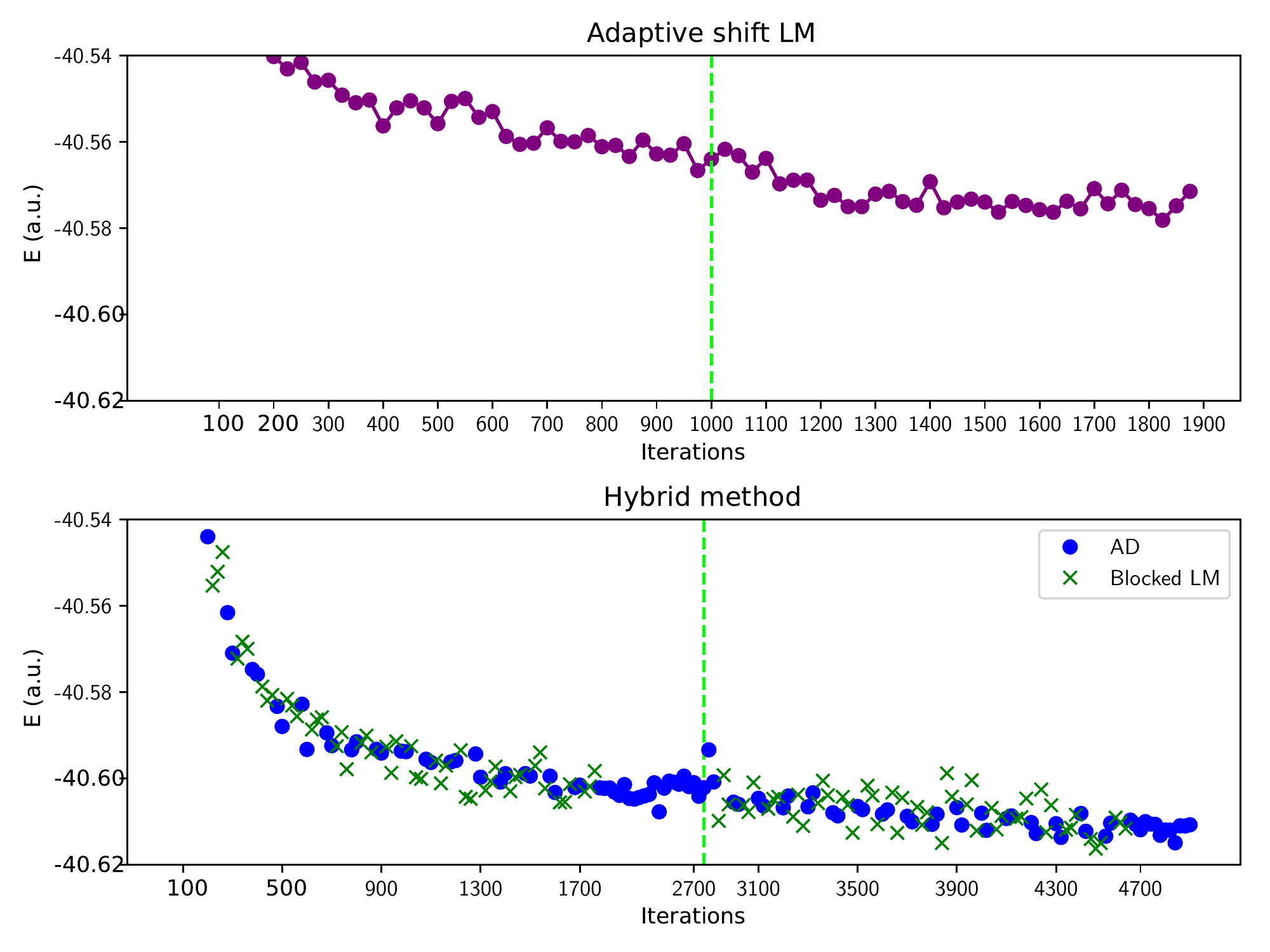} 
\caption{Zoomed in view of excited state energy for adaptive shift LM and hybrid method. Green dashed line 
marks the end of the first stage where only Jastrow and CI parameters are optimized.}
\label{fig:zoomed_energy}

\end{figure}

\begin{figure}[H]

\includegraphics[width=\textwidth]{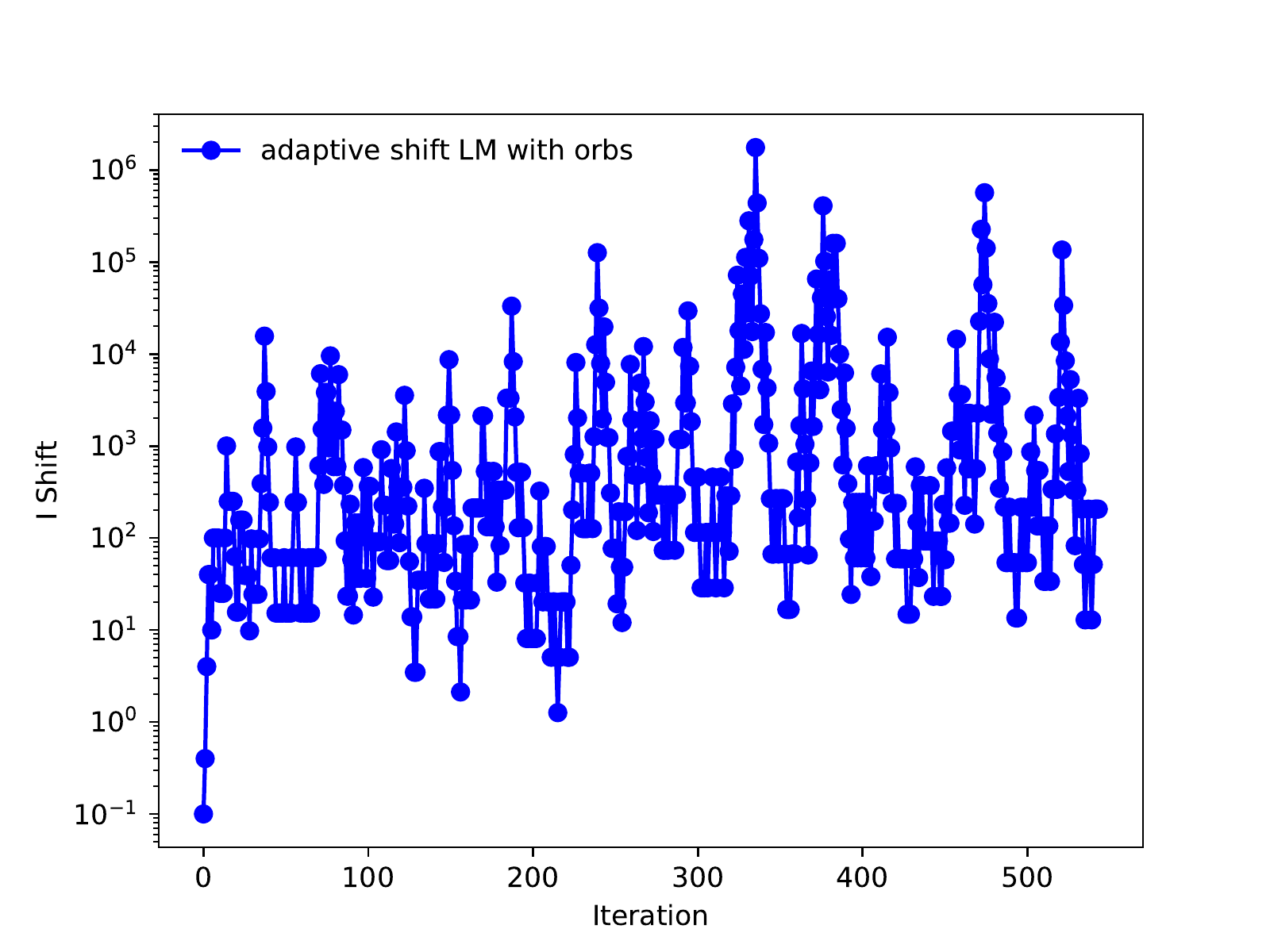} 
\caption{Value of the diagonal shift $c_I$ for the adaptive shift LM during the orbital optimization stage. The overlap shift $c_S$ is a factor of 10 larger than $c_I$ and is adjusted in tandem.} 
\label{fig:adapt_lm_shift}

\end{figure}

The change from fixed to adaptive shifts yields far better performance for the LM, with 
steady decreases in objective function, energy, and variance.
We provide a zoomed in plot of the energy for the adaptive shift LM and hybrid method 
in Figure \ref{fig:zoomed_energy} to make clear that both approaches continue to 
lower the energy after enabling orbital optimization.
The adaptive shift LM optimization makes only rather slow progress over many 
hundreds of iterations and after a few hundred iterations, as depicted in
Figure \ref{fig:adapt_lm_shift}, the 
shifts are so large that only very minor changes are being made to the parameters in the 
remainder of the calculation.
While this sluggish optimization would be undesirable in normal VMC applications, it 
clearly does not show any pathological behavior in terms of instability even though 
the adaptive shift LM is operating with only 100,000 samples 
per iteration.
Although this low number of samples may hamper the LM's ability to determine 
effective parameter updates that would minimize more quickly, any stability challenges 
it might pose have been overcome with enhanced step control in the algorithm's design.
Essentially, the adaptive shifts prevent the optimizer from pushing past the point where statistical 
uncertainty precludes effective update steps.
While the adaptive shift LM's optimization progress is slowed dramatically by the growth of the shifts, we note that the hybrid method does not face this possibility due to resetting the shifts after each macro-iteration. We also emphasize that whether the LM is able to reach the objective function minimum within a reasonable amount of optimization effort is a different issue from whether its 
optimization is proceeding stably.
The hybrid method is similarly stable and able to achieve significantly lower values of $\Omega$, energy, and variance.
The blocked LM iterations show some scatter in the objective 
function and the variance, but do not derail the stability of the optimization.
The separation between the blocked LM and AD variance points in Figure \ref{fig:all_methods_variance}, also seen in Figure  \ref{fig:lm_all_large_cas_vary_omega}, is an autocorrelation 
artifact from the low number of samples used to evaluate the AD variances being averaged 
together in each point and diminishes if AD iterations are given more samples (we do not do 
so as it is not necessary for effective optimization).

\begin{figure}[H]

\includegraphics[width=\textwidth]{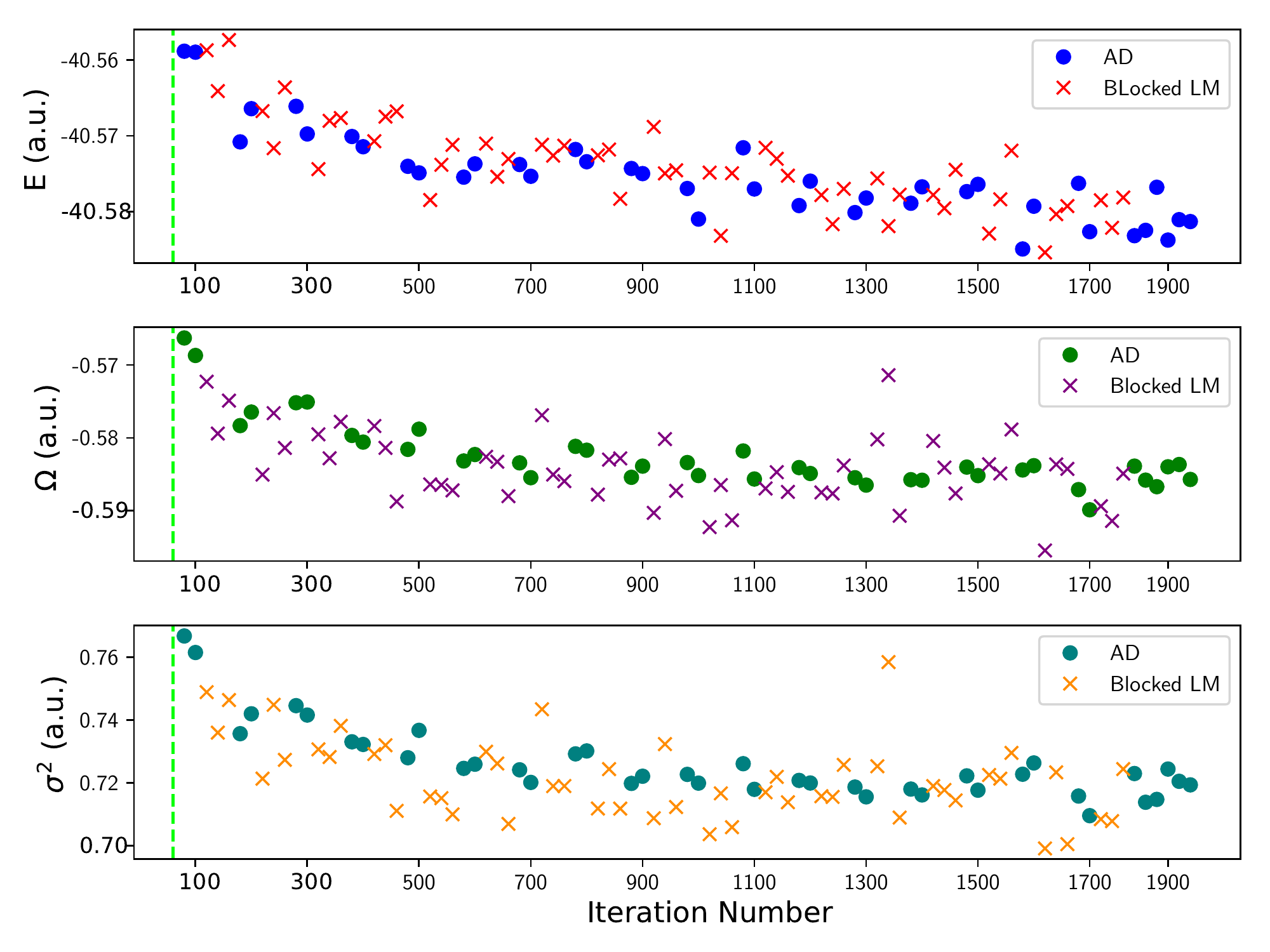} 
\caption{Optimization of all parameters including orbitals using the hybrid method 
starting from the end of the LM fixed shift optimization of only the Jastrow and CI parameters.
Green line is shown to the left to make clear these plots show only the stage with orbital optimization.}
\label{fig:lm_fixed_shift_start_hybrid_opt}

\end{figure}

We also consider the potential effects of the optimized MSJ starting point on the second, 
orbital optimizing, stage of the optimization.
As depicted earlier in Figure \ref{fig:all_methods_target}, the hybrid method 
achieves a lower value of $\Omega$ than the two types of the LM during the first Jastrow 
and CI phase of optimization 
and therefore has a better initial wave function when orbital optimization is turned on.
To check whether this difference has any stability consequences, we also performed a 
hybrid method optimization of all parameters beginning from the optimized MSJ ansatz 
obtained by the fixed shift LM.
As shown in Figure \ref{fig:lm_fixed_shift_start_hybrid_opt}, the hybrid method is 
still able to stably optimize from this inferior starting point.
While the quality of the initial wave function may still influence how fully the hybrid 
method is able to minimize $\Omega$, a somewhat poorer starting point does not 
destabilize the hybrid algorithm.

\begin{figure}[H]

\includegraphics[width=\textwidth]{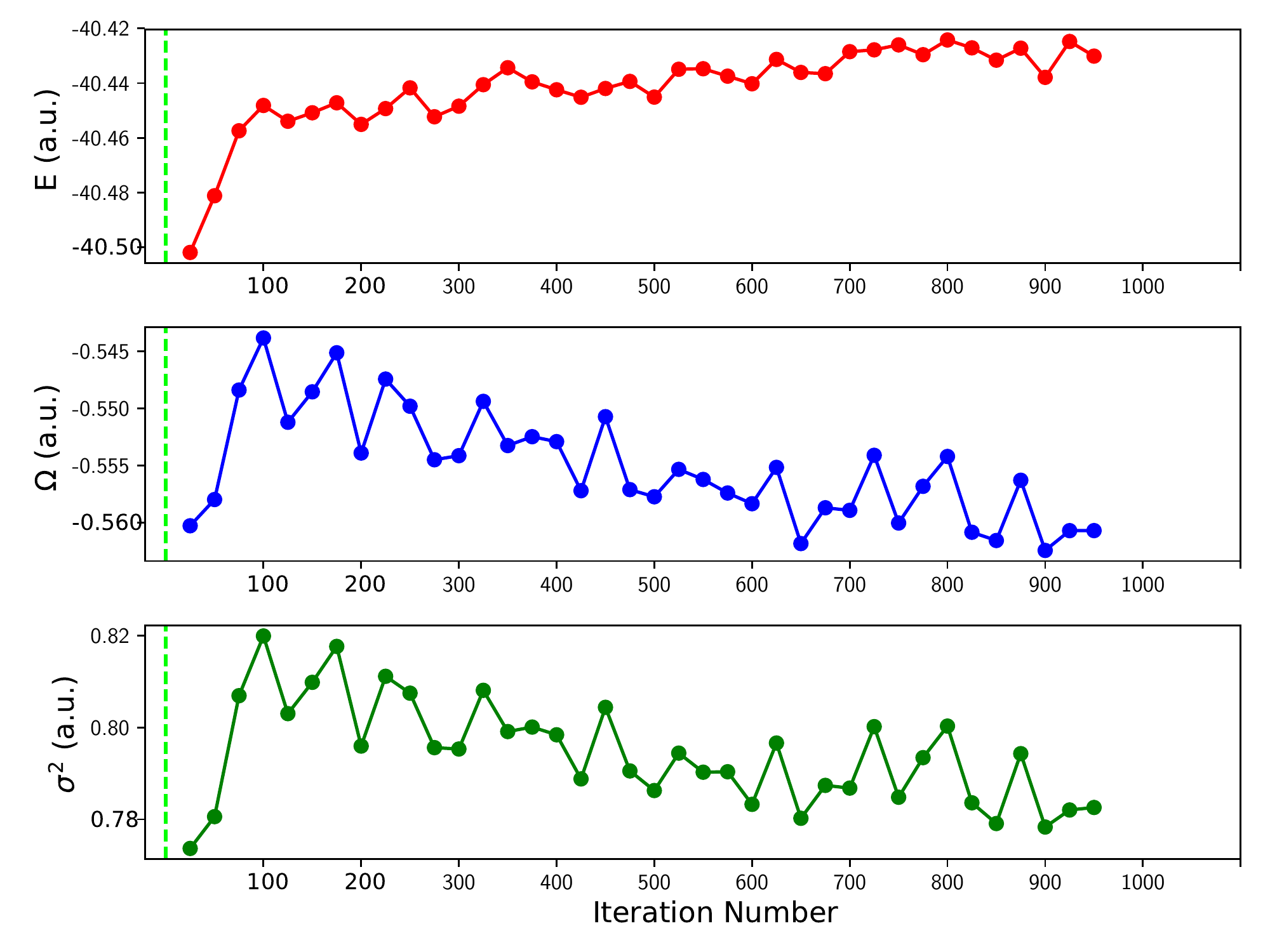} 
\caption{Optimization of all parameters including orbitals using the LM with fixed shifts while varying the energy targeting parameter $\omega$ on the fly. Starting point was the optimized MSJ wave function from the preceding LM fixed shift optimization of only the Jastrow and CI parameters. Green line is shown to the left to make clear these plots show only the stage with orbital optimization.}
\label{fig:lm_all_large_cas_vary_omega}

\end{figure}

\begin{figure}[H]

\includegraphics[width=\textwidth]{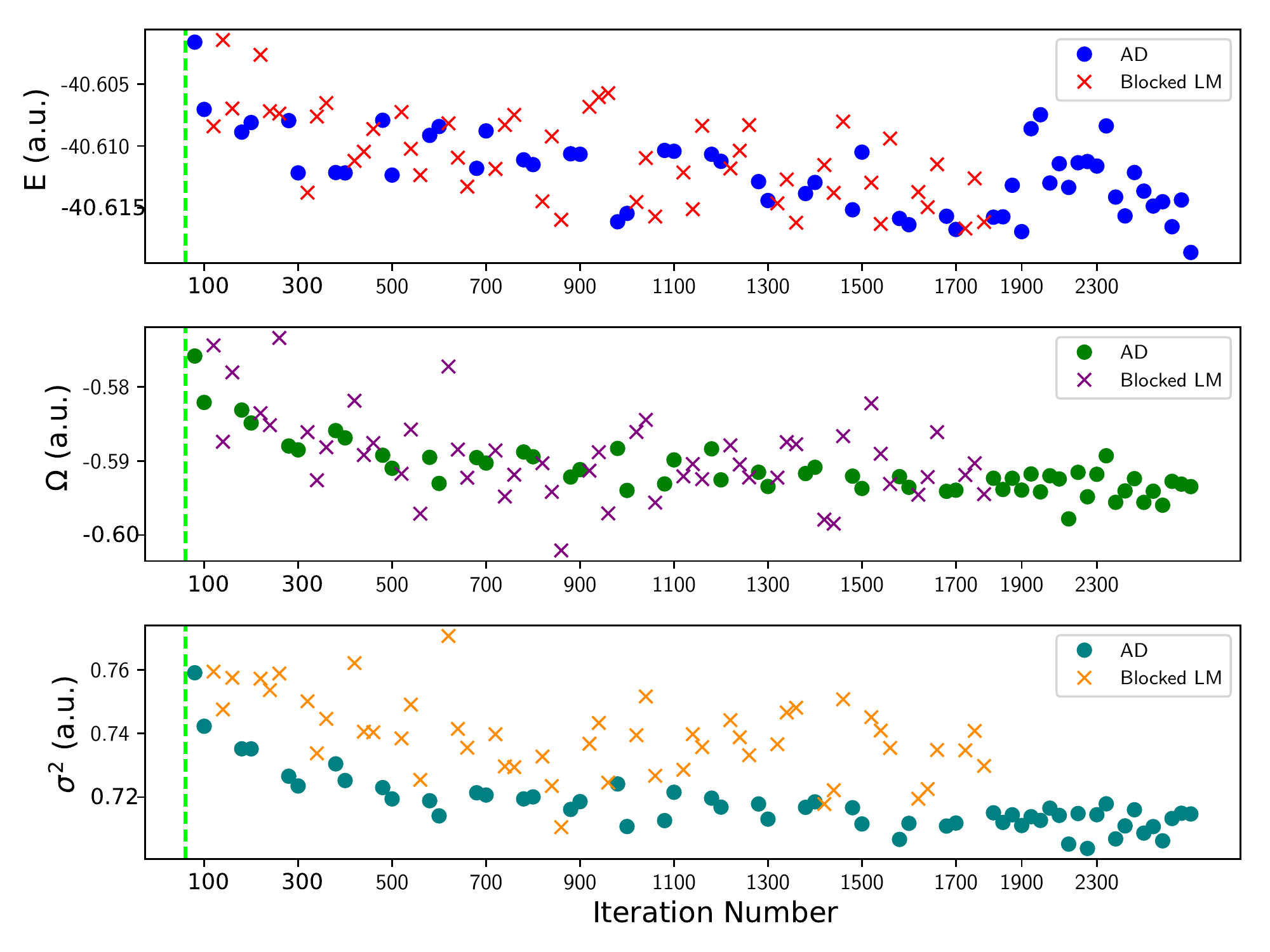} 
\caption{Optimization of all parameters including orbitals using the hybrid method while varying $\omega$. Starting point was the optimized MSJ wave function from the preceding hybrid method optimization of only the Jastrow and CI parameters. Green line is shown to the left to make clear these plots show only the stage with orbital optimization. The separation between AD and blocked LM variance points is an artifact of autocorrelation.}
\label{fig:hybrid_all_large_cas_vary_omega}

\end{figure}

\begin{figure}[H]

\includegraphics[width=\textwidth]{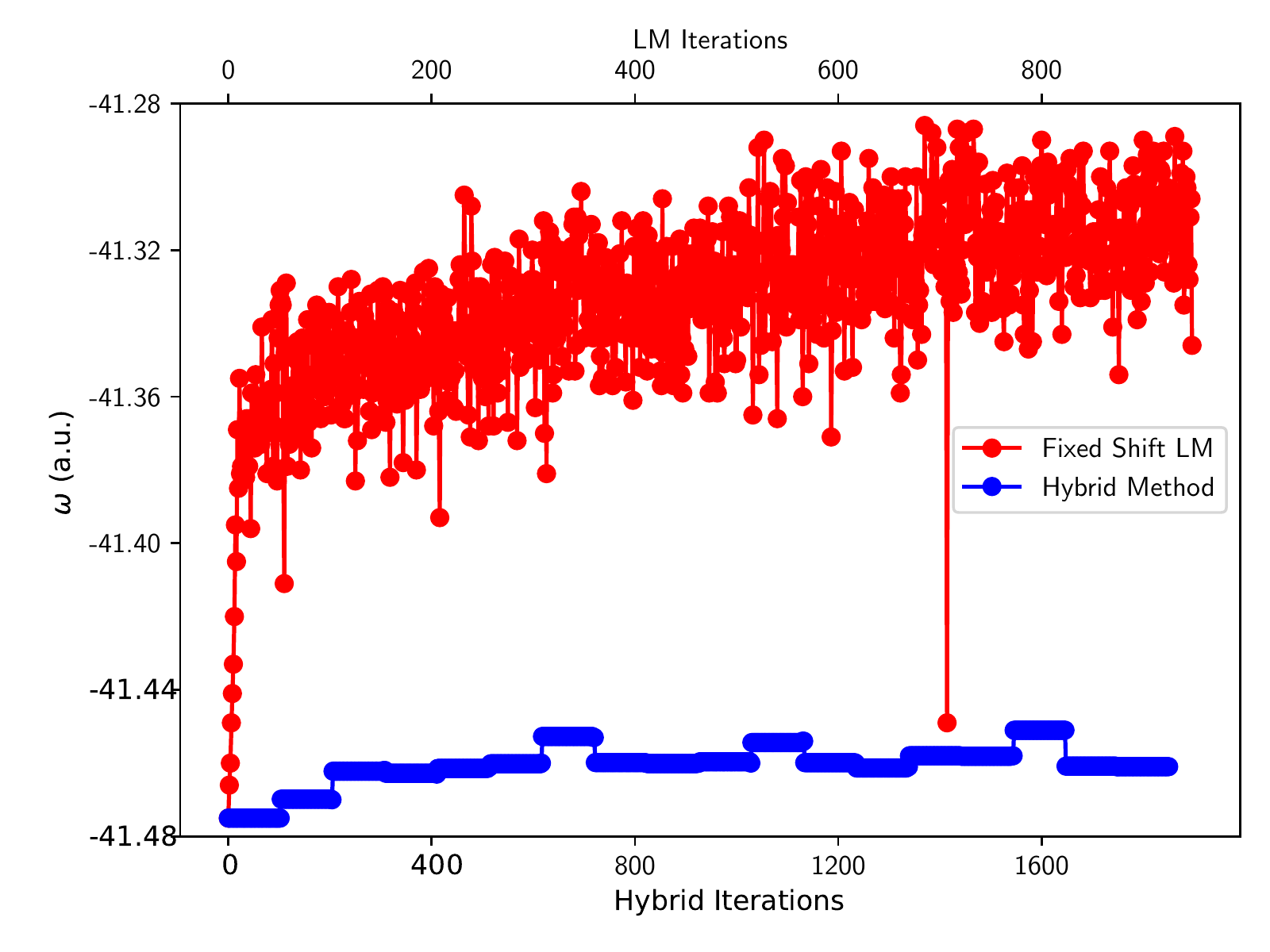} 
\caption{Varying values of omega over individual iterations during the fixed shift LM and hybrid method optimizations 
of all parameters depicted in Figures 11 and 12. The last 1000 iterations of AD in the hybrid optimization used the final value of omega on the last hybrid macro-iteration and are omitted from this plot.}
\label{fig:omega_variations}

\end{figure}

The optimizations shown so far have all used a fixed value of $\omega$.
While in practical calculations, a series of fixed $\omega$ calculations is always an option 
for moving $\omega$ to transform minimization of $\Omega$ into variance minimization, for 
completeness we also consider varying $\omega$ over the course of a single 
optimization.
For the fixed shift LM and the hybrid method, we again turn on orbital optimization at 
their respective optimized Jastrow and CI starting points, now with schemes for varying $\omega$.
For the LM, during the first 10 iterations, we interpolate\cite{Shea2017} $\omega$ between its initial value and $E - \sigma$, and afterward allow $\omega$ to float at the value 
of $E - \sigma$ set by the preceding iteration.
For the hybrid method, we use the last 30 AD iterations in each macro-iteration to compute
$E -\sigma$ for setting a new value of $\omega$ for the blocked LM steps and the AD 
section of the next macro-iteration.
As seen in Figures \ref{fig:lm_all_large_cas_vary_omega} and \ref{fig:hybrid_all_large_cas_vary_omega}, there is little difference in the methods' 
stability behavior from the fixed $\omega$ case.
The fixed shift LM again shows a clear rise in the energy and variance while the hybrid method 
still performs a stable minimization.
In Figure \ref{fig:lm_all_large_cas_vary_omega}, the objective function $\Omega$ and the 
variance mostly recover while the energy does not, suggesting the lack of adaptive step 
control may be allowing the LM optimization to move into another basin of 
convergence.
As Figure \ref{fig:omega_variations} shows, the value of $\omega$ is dragged up 
significantly during the unstable fixed shift LM optimization, while it only moves a little during the hybrid method.
The hybrid method's stability suggests that varying $\omega$ and thus changing the objective 
function landscape on the fly are not the key factors in determining stability.
Instead, we see again that the presence or absence of adaptive step control makes the 
difference.

\begin{figure}[H]

\includegraphics[width=\textwidth]{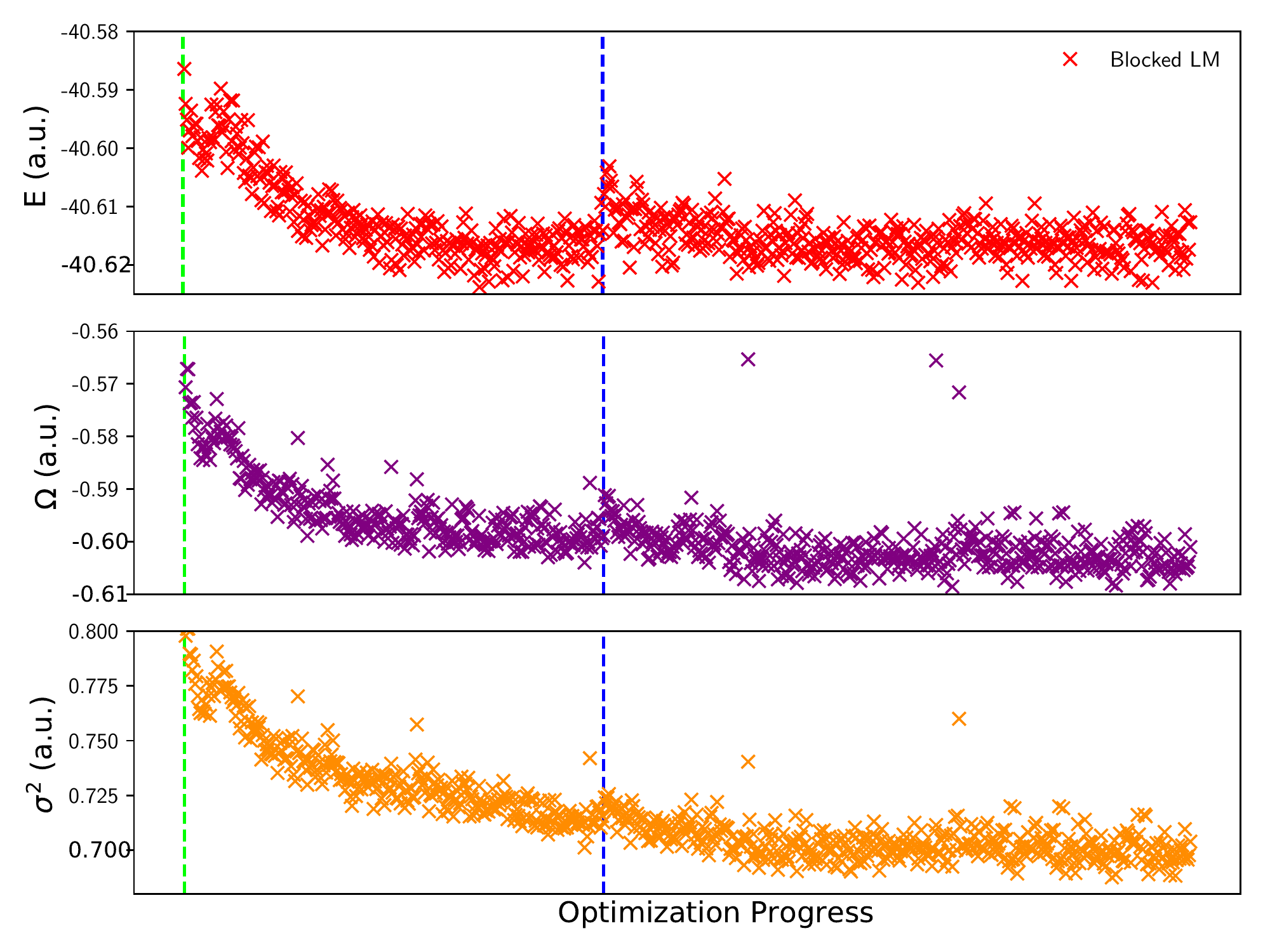} 
\caption{Hybrid method optimization of a larger excited state ansatz with 7168 determinants. Green line is shown to the left to make clear these plots show only the stage with orbital optimization. Blue line marks the beginning of varying $\omega$. A total of 744 blocked LM and 31,200 AD iterations were used (309 blocked LM and 15,200 AD for the fixed omega phase and the remainder with varying omega). For clarity, we show only the blocked LM iterations. AD used 40,000 samples per iteration while the blocked LM used 1,000,000. The blocked LM divided the parameters among 30 blocks, retaining 30 parameter directions per block along with 5 directions from AD. } 
\label{fig:full_hybrid_opt}

\end{figure}

We finally consider a larger ansatz and longer optimization run as an even more stringent test of the stability of the adaptive step controlled hybrid method.
We again employ the (6e,10o) CAS, now enforcing symmetry to obtain a state-specific CASSCF description of the excited state and taking all 7168 CAS determinants.
We again use a cc-pVDZ 
basis set with BFD pseudopotentials.\cite{Burkatzki2007}
This wave function should be nearly identical to one of the ansatzes for which Filippi and coworkers saw instabilities
(see Figure 3 of Ref. \citenum{Cuzzocrea2020}), but may still differ in some details such as the parameterization of the Jastrow factor.
We optimize using the hybrid method and again follow a staged strategy of first optimizing Jastrow and CI parameters before turning on orbital optimization.
For this more challenging optimization, we use one million samples per blocked LM iteration and divide parameters into 30 blocks, but otherwise use the same hybrid method settings as described for the earlier stability tests shown in Figures \ref{fig:all_methods_energy}-\ref{fig:all_methods_variance}.

The final stage of this longer optimization, in which all parameters including orbital parameters are optimized, has a total of 11440 variational parameters and is depicted in Figure \ref{fig:full_hybrid_opt}, where for clarity we show only the blocked LM iterations of the hybrid method.
We consider both optimization with a fixed value of $\omega = -41.475$ and then allowing $\omega$ 
to vary to verify stability in both cases.
During the varying phase, $\omega$ is updated after every 9 blocked LM iterations to the median 
value of $E- \sigma$ from those 9 preceding iterations.
We see in Figure \ref{fig:full_hybrid_opt} that the energy and variance can be fully converged.
For assessing computational cost, we emphasize that the AD iterations individually require far 
less sampling effort than the blocked LM steps.
The combined sampling effort of the total 744 blocked LM iterations and 31,200 AD iterations is 
equivalent to 1992 blocked LM iterations alone, which is only about two and a half times as many iterations as 
used in previously reported Newton optimizations.\cite{Cuzzocrea2020}
A higher number of iterations is expected in general for optimizers with step control and
we note that in practical calculations, the varying of $\omega$ could be begun earlier to obtain 
the final variance minimization in fewer iterations.

As in the stability tests that involved shorter runs, we observe no signs of instabilities.
This stands in marked contrast to previous optimizations on
similarly large multi-Slater expansions that
employed Newton's method without adaptive step control,
where instabilities were clearly visible within the first
100 optimization steps (see Figures 3, S2, S3 and S4 of
Ref.\ \citenum{Cuzzocrea2020}).
We also note that the optimization in Figure\ \ref{fig:full_hybrid_opt} achieves a lower variance (0.70 a.u.)
than that achieved by the previously reported
\cite{Cuzzocrea2020} unstable Newton
optimization (0.73 a.u.).
This may in part be due to differences in the Jastrow factor
details, but the fact that this lower variance is
now achieved in tandem with a stable optimization
is significant.
Of course, our energy is higher than the value reported for energy minimization of the CAS(6,10) 
ansatz,\cite{Cuzzocrea2020} which is the expected outcome whenever variance and energy 
minimizations are compared for an approximate ansatz and does not by itself imply anything 
about optimization stability.
With this demonstration in a larger wave function and longer optimization run, we have additional reassurance that the hybrid optimizer with adaptive step control can yield stable variance-based optimization.

\section{Conclusions}

The sequences of optimizations we have performed point to several considerations when seeking
to ensure stable and successful applications of variance-based state-specific VMC.
Scenarios characterized by large numbers of difficult parameters, as in orbital 
optimization, pose a significant risk of optimization instability from poor parameter 
updates.
This difficulty is exacerbated if the amount of sampling is insufficient to overcome 
uncertainty coupling in the matrix diagonalization of large parameter sets or to be able to 
recognize and reject bad updates.
Our findings on stability failures of the LM add to other observations of challenges in 
variance minimization\cite{Cuzzocrea2020} and help clarify when such optimization 
difficulties are most likely to be encountered.
The nature of the optimization failures we encounter point to choices in optimizer design as 
the driver of variance minimization instabilities and do not offer any indication that shape 
of the variance surface is pathological.
While the problem of optimization instability will remain a concern as researchers 
pursue studies of larger systems and more complex ansatzes, multiple tools are available 
for curbing its impact.

First, the ability to adaptively control step size and direction is a key element of 
robustness in challenging optimizations.
Previously reported instabilities were seen in a setting that lacked adaptive step control, as were 
all the unstable runs in the present study.
In contrast, optimizations in which we employ adaptive step control (either in the LM alone or in the
hybrid method) proved stable.
Second, algorithms can be designed to reduce the optimization burden on the portions 
most prone to instability.
This idea is reflected in the hybrid method's structure, with a large portion of the 
optimization left to the more stable AD and the reduction of the 
size of the LM matrices by first including only parameters of significant gradients and 
then blocking.
Finally, careful construction of wave functions to effectively describe correlation 
while remaining readily optimizable at the VMC level is advantageous.
We have shown that determining orbital shapes with state-specific quantum chemistry and 
optimizing only Jastrow and CI parameters is sufficient for an accurate excitation energy in 
CN5 and demonstrated the same idea for a broader set of molecules in recent 
work.\cite{Otis2021}
Circumventing VMC orbital optimization in this way avoids any stability challenges it 
might pose and greatly enhances computational efficiency.
Taken together, these approaches provide reassurance that state-specific variance-based VMC 
can be performed in a robust and stable manner so long as adaptive step control is employed.

\section{Acknowledgements}
This work was supported by the Office of Science, Office of Basic Energy Sciences, the U.S. 
Department of Energy, Contract No. DE-AC02-05CH11231.
Computational work was performed with the Berkeley Research Computing
Savio cluster and the LBNL Lawrencium cluster.

\section{Disclosure Statement}
No potential conflict of interest was reported by the authors.

\section{Appendix}

  \begin{table}[H]
%\small
%\footnotesize
     \caption{Structure of CN5. Coordinates in \r{A}.}
     \centering
     \begin{tabular}{lSSS}
       \multicolumn{1}{l}{} &  &  & \\ \hline
C   &   0.000000  &  0.000000  &  0.340620  \\
C   &   0.000000  &  1.188120  & -0.363814  \\
C   &   0.000000  & -1.188120  & -0.363814  \\
H   &   0.000000  &  0.000000  &  1.424530  \\
H   &   0.000000  &  1.160200  & -1.449150  \\
H   &   0.000000  & -1.160200  & -1.449150  \\
N   &   0.000000  &  2.389120  &  0.164413  \\
N   &   0.000000  & -2.389120  &  0.164413  \\
H   &   0.000000  &  2.531430  &  1.161940  \\
H   &   0.000000  & -2.531430  &  1.161940  \\
H   &   0.000000  &  3.211380  & -0.414923  \\
H   &   0.000000  & -3.211380  & -0.414923  \\
       \hline
     \end{tabular}
     \label{tab:ch2sGeo}
   \end{table}

%\bibliographystyle{tfo}
%\bibliography{interacttfosample}
\printbibliography

\clearpage

\begin{figure}[H]

\includegraphics[width=\textwidth]{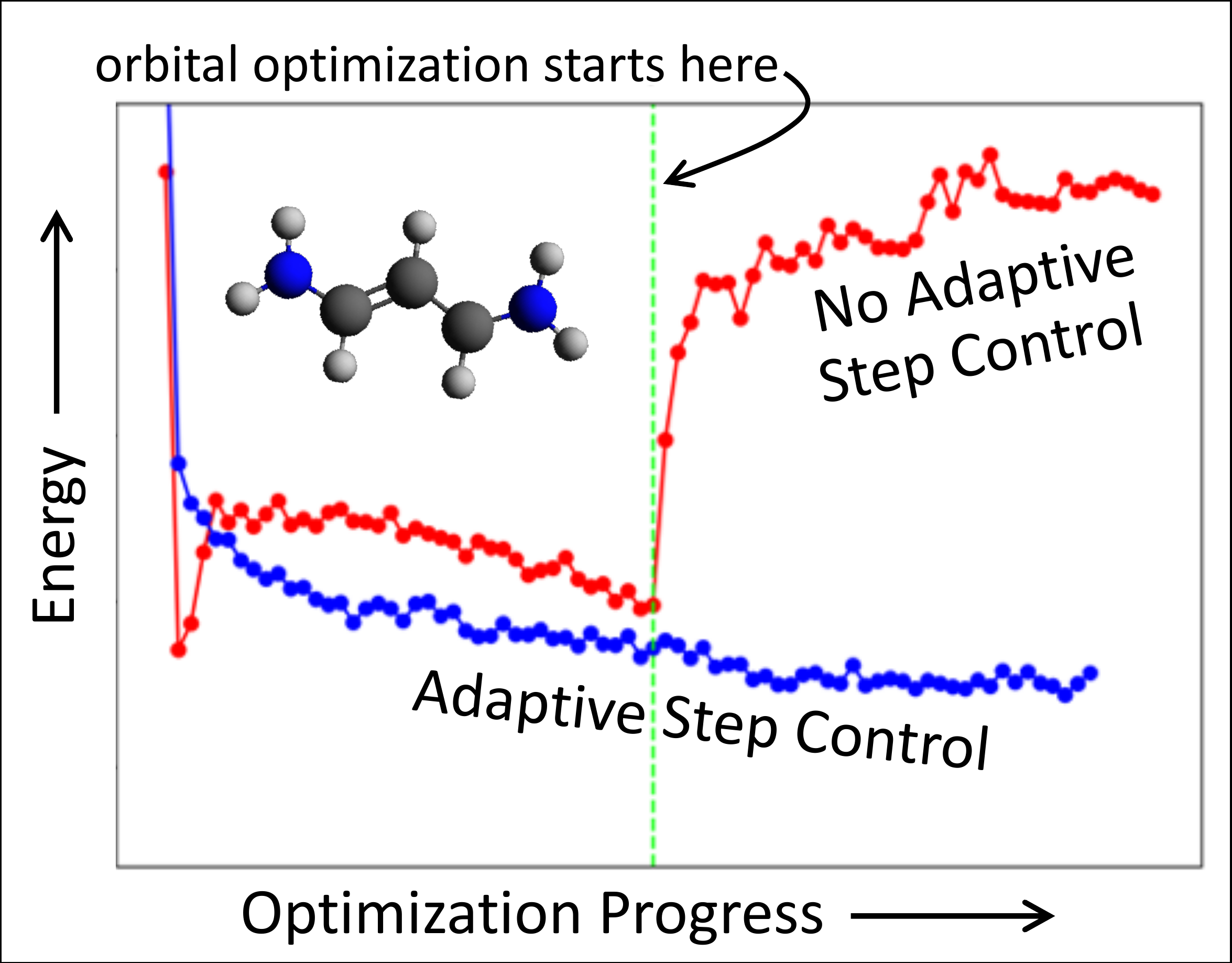} 
\caption{For Table of Contents Only}
\label{fig:toc_image}

\end{figure}

\end{document}